\documentclass{sig-alternate-10pt}
\usepackage{epsfig,endnotes}

\usepackage{flushend}

\usepackage{setspace}
\usepackage[numbers,sort&compress]{natbib}
\usepackage{xspace}
\usepackage{graphicx}
\usepackage{thumbpdf}
\usepackage{url}
\usepackage{amsmath}
\usepackage{subfig}

\newcommand{\mytit}[0]{\textsc{BaPu}\xspace}
\newcommand{\basic}[0]{\textsc{BaPu}\xspace}
\newcommand{\proack}[0]{\textsc{BaPu-Pro}\xspace}
\newcommand{\buffering}[0]{\textsc{SimpleBuffer}\xspace}

\newcommand{\ap}[0]{\emph{\mytit-AP}\xspace}
\newcommand{\gw}[0]{\emph{\mytit-Gateway}\xspace}
\newcommand{\s}[0]{\emph{Sen\-der}\xspace}
\renewcommand{\d}[0]{\emph{Receiver}\xspace}
\newcommand{\home}[0]{\emph{Home-AP}\xspace}
\newcommand{\mon}[0]{\emph{Monitor-AP}\xspace}

\newcommand{\fixme}[1]{{\bf FIXME: #1}}

\newcommand{\Hide}[1]{}
\let\ignore\Hide

\begin{document}

%don't want date printed
\date{}

%make title bold and 14 pt font (Latex default is non-bold, 16 pt)
\title{{\huge \sc \mytit}: Efficient and Practical Bunching of\\ Access Point Uplinks}

%for single author (just remove % characters)
\author{
Tao Jin \quad Triet Vo-Huu \quad Erik-Oliver Blass \quad Guevara Noubir\\
       \affaddr{Northeastern University}\\
       \affaddr{Boston, MA, USA}\\
       \email{\{taojin|vohuudtr|blass|noubir\}@ccs.neu.edu}
%\and
%{\rm Triet Vo-Huu}\\%
%Northeastern University
%\and
%{\rm Erik-Oliver Blass}\\
%Northeastern University
%\and
%{\rm Guevara Noubir}\\
%Northeastern University
} % end author

\maketitle

% Use the following at camera-ready time to suppress page numbers.
% Comment it out when you first submit the paper for review.
%\thispagestyle{empty}  

\subsection*{Abstract}

Today's increasing demand for wirelessly uploading a large volume of User Generated Content (UGC) is still significantly limited by the throttled backhaul 
of residential broadband (typically between 1 and 3Mbps). 
We propose \mytit, a carefully designed system with implementation for bunching WiFi access points' 
backhaul to achieve a high aggregated throughput. \mytit is inspired by a decade of networking design principles and 
techniques to enable efficient TCP over wireless links and multipath. 
\mytit aims to achieve two major goals:
1) requires \emph{no client modification} for easy incremental adoption;
2) supports \emph{not only} UDP, but also TCP traffic to greatly extend its applicability to a broad class of popular applications such as HD streaming or large file transfer. We prototyped \mytit with commodity hardware. Our extensive experiments shows that despite TCP's sensitivity to typical channel factors such as high wireless packet loss, out-of-order packets arrivals due to multipath, heterogeneous backhaul capacity, and dynamic delays, \mytit achieves a backhaul aggregation up to 95\% of the theoretical maximum throughput for UDP and 88\% for TCP. We also empirically estimate the potential idle bandwidth that can be harnessed from residential broadband.

\section{Introduction}
\label{sec:introduction}

Nowadays, the mobile devices are equipped with high-resolution cameras 
and a variety of sensors and are quickly becoming the primary device to 
generate personal multimedia content. Both the quality and quantity of User
Generated Content grows continuously.  This  naturally leads to end users' ever
increasing demand of sharing these high volume of UGC with others 
in an instant way. Prominent examples of services allowing multimedia content sharing are YouTube, Dailymotion, and various social networking platforms 
like Facebook and Google+. In addition, there is also a trend of instantly 
backing up personal files in the ``Cloud Storage", such as Dropbox and iCloud.
To obtain a satisfactory user experience, users need sufficient uplink bandwidth to 
do the fast bulk data transfer to the Internet. However, today's ISP's generally offer
highly throttled uplink bandwidth around 1 to 3Mbps.  
As a result, instant sharing of HD content or fast data backup in the ``Cloud"
is generally infeasible in today's residential broadband. 
For example, iPhone 5 takes video at 1080p and 30fps, which translates to 
around 200MB per minute.  With 3Mbps uplink, it takes over an hour to
upload a 10 minute video clip to iCloud!  These limitations are
even more critical for users who desire to retain the control over
their content and intend to share them directly from their homes.
This calls for solutions to scale backhaul uplink.

\begin{figure}
\centering
\includegraphics[width=0.75\columnwidth]{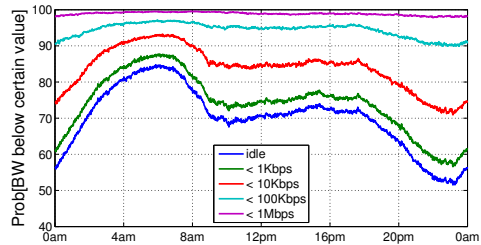}
\vspace{-7pt}
\caption{\label{fig:idle} CDF of uplink bandwidth usage (per household) in residential broadband.}
\end{figure}

\begin{figure}
\centering
\includegraphics[width=0.75\columnwidth]{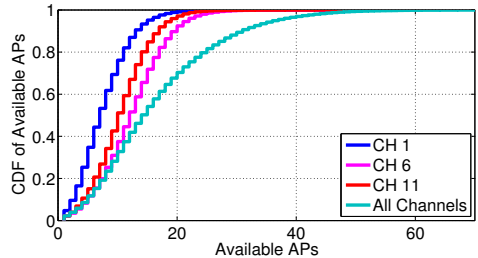}
\vspace{-7pt}
\caption{\label{fig:wardriving_ap_density_per_channel} 
Available APs per scanning in Wardriving.}
\end{figure}

In this work, we propose a complete software based solution on WiFi Access Point for aggregating multiple broadband uplinks, with the assistance of the WiFi infrastructure in the same neighborhood.  Our solution features complete transparency to client devices and high aggregated throughput for both TCP and UDP, even in lossy wireless environment.  Our work is primarily motivated by the following observations: 
\begin{itemize}
  \setlength{\itemsep}{0pt}
  \setlength{\parskip}{0pt}
  \setlength{\parsep}{0pt}
\item {\bf Asymmetric WiFi and broadband capacity:} In
  contrast to the broadband uplink, WiFi has a much higher bandwidth. 802.11n
  supports up to 600Mbps data rate. With sufficiently high WiFi
  bandwidth, it is beneficial to wirelessly communicate with multiple proximate APs 
  and ``harness" the idle broadband uplink bandwidth behind them.

\item {\bf Mostly idle broadband uplinks}: Since February 2011, we have
  developed and deployed a WiFi testbed \cite{open-infrastructure} in Boston 
  urban area, aiming to monitor the usage pattern of residential broadband networks.    
  As shown in Table \ref{table:hbt_summary}, this testbed consists of 30 home WiFi 
  APs running customized firmware based on OpenWRT \cite{openwrt}.  
  Each AP reports the network statistics 
  every 10 second.  During a 18 month period, we have collected over 70 million 
  records. We observe that the broadband uplink utilization is very low. Figure 
  \ref{fig:idle} shows the probability of uplink bandwidth being consumed at most 
  certain value during a 24 hour time window.  Throughout the day, there is at least 
  50\% chance that uplink is completely idle. Even
  during peak hours, there is over 90\% chance that the uplink
  bandwidth usage is below 100Kbps.  This implies that there exists a considerable 
  amount of idle uplink bandwidth resources, which makes bandwidth harnessing
  through neighboring APs a viable approach for scaling the uplink
  capacity.
  
\begin{table}
\centering
\begin{tabular}{lc}
\hline
\textbf{Location}&Boston urban area\\
\textbf{Home APs}&Comcast (26), RCN (4)\\
\textbf{Data collection time}& Feb. 2011 $\sim$ Dec. 2012\\%now\\
\textbf{Network stats samples}& 70 million\\
\hline
\end{tabular}
\caption{Data summary of Broadband usage statistics collected from residential WiFi testbed}
\label{table:hbt_summary}
\end{table}

\item {\bf WiFi densely deployed in residential area:} The density
  of WiFi APs is very high in residential areas. Already in 2005, authors in~\cite{AkellaJSS2007} measured more than 10 APs per geographical
  location.  Recently, we conducted Wardriving measurements 
  in 4 urban residential areas in Boston. Our results (Table \ref{table:wardrv_summary})
   indicate 3192 unencrypted WiFi APs, accounting for 14.2\% of total APs detected during our wardriving.  As shown in Figure \ref{fig:wardriving_ap_density_per_channel}, there are on average 17 APs available 
  at each location, with on average 7 to 12 APs on each channel. This enormous
  presence of WiFi APs also justifies the feasibility of the concept of bandwidth
  aggregation via open APs. 

\begin{table}[ht]
\centering
\begin{tabular}{lc}
\hline
\textbf{Total APs}&22,475 (100\%)\\
\textbf{Unencrypted APs}& 3,192 (14.2\%)\\
\hline
\end{tabular}
\caption{Boston Wardriving Data Summary}
\label{table:wardrv_summary}
\end{table}
\vspace{-10pt}

\item {\bf WiFi becoming open and social:}
Nowadays, end users have an ever increasing demand of ubiquitous Internet access. Driven by such demand, there is an emerging model that broadband subscribers host two WiFi SSIDs, one encrypted for private use, the other unencrypted to share part of their bandwidth as public WiFi signal to mobile users for free or for some payment in return.  Unencrypted guest-WLAN is now a standard feature in mainstream home WiFi AP products like LinkSys and D-Link.  FON \cite{fon}, a leading company in this area, claims to have over 7 million hotspots worldwide.  In addition, WiFi APs are quickly becoming cloud-managed devices, such as FON and Meraki \cite{meraki}.  AP firmware updates are regularly pushed from the Cloud.  Given such trend of WiFi quickly becoming social and cloud-powered, we believe a software based solution on WiFi AP can make a quite easy incremental adoption of our technology.

\item {\bf Lack of efficient and practical solution:} Despite a set of prior work 
exploring how to aggregate wired bandwidth through WiFi, they either require heavy modification on client, or support only specific application, such as UDP based bulk file transfer~\cite{link-alike}.  Our goal is to design a client transparent, software based solution, which is easy to deploy and offer generic support for both TCP and UDP applications.
\end{itemize}

With the above motivations and goals bearing in mind, we design our \mytit system.  Our major contributions in this work are summarized as follows: 
\vskip 1eX

\textbf{Transparency to client}: \mytit does not require any modification to client devices.  The client device, running in Station mode, transfers data via unicast link to its ``home'' AP.  Given the broadcast nature of wireless communication, such unicast packets can be ``heard'' by both ``home'' AP and the ``neighboring'' APs on the same channel.  They each upload a share of 
received (overheard) packets to the destination in a collaborative manner.  Such 
transparency to client devices allows all kinds of wireless client devices and a broad class of legacy network applications, such as streaming and large file transfer, to seamlessly utilize \mytit system.

\textbf{Efficient aggregation for both TCP and UDP}:
Given our design goal of client transparency, some commonly adopted technique in the existing bandwidth aggregation solutions, such as parallel TCP flows \cite{KandulaLBK2008}, are no longer valid, because it requires client applications to intentionally establish multiple connections through different APs and transfer data in parallel.  The multiplexing of a single TCP flow through multiple paths raises many technical challenges which makes efficient aggregation non-trivial. Our initial approach relied on coding across paths, however, we could show that a conceptually simpler mechanism, which we call \emph{Proactive-ACK}, combined with a reliable 802.11 unicast to the ``home'' AP, and adequate scheduling are sufficient.
%{\bf FIXME: wording} We propose a mechanism called \textit{\bf Proactive-ACK}, and empirically show its high aggregated throughput.  Besides, we use a set of experiment data to justify our design choice. 

\textbf{Prototype with commodity hardware}: We  prototyped our
complete \mytit system on commodity WiFi APs.  We flash Buffalo 802.11n APs with Linux based OpenWRT~\cite{openwrt} firmware.  As of today, OpenWRT supports devices by more than 50 manufacturers and hundreds of models.  This gives us a great selection of compatible devices to deploy \mytit. 

\textbf{Evaluation}: We have conducted an extensive set of experiments to evaluate \mytit in various realistic network settings.  Our results show that \mytit achieves high
aggregated throughput in UDP transmission, harnessing over 95\% of total uplink bandwidth.  With the \emph{Proactive-ACK} mechanism, \mytit harnesses over 88\% of total uplink bandwidth in TCP transmission.
%  {\bf FIXME: should we mention comparison with existing solution?} In comparison with existing solution, BaPu significantly increases the performance by over 60\%.

\textbf{Design guideline for bandwidth sharing}: We propose a simple traffic shaping method on AP, which allows us to harness as much idle bandwidth as possible without affecting home users' regular network usage.  We also give an estimation of idle uplink bandwidth under a typical residential broadband usage pattern.

Our paper is organized as follows. We first present an overview of \mytit system.\ignore{ along with
two typical application scenarios.}  The details of our design is discussed in
Section~\ref{sec:bapu}.  We evaluate the performance of \mytit in Section~\ref{sec:evaluation}. In Section~\ref{sec:uplink}, we quantitatively evaluate the potential impact of uplink sharing to home users' regular network usage. We discuss related work in Section~\ref{sec:related} and conclude the paper in Section~\ref{sec:conclusion}.
 % {\bf FIXME: maybe need to add a new feasibility study section for uplink bandwidth sharing}

%BaPu does not require any modification
%to client devices. BaPu runs on home WiFi APs or online services' gateways, 
%and multiplexes legacy clients' WiFi traffic through multiple backhaul links 
%transparently. 

\ignore{
There exists a plethora of previous research that considers
cooperation of APs to scale the quality and capacity of wired
backhauls, cf.~\citet{GiustinianoGLR2009, GiustinianoGTLDMR2010,
  fatvap, NicholsonWN2009}.  Yet, previous research focuses primarily
on using one physical WiFi card seamlessly to switch among multiple
APs in order to harness idle bandwidth -- while ascertaining fairness
among APs.  However, such solutions may boost only \emph{download}
throughput.  Upload of a TCP or UDP flow is usually assigned to a
single AP, therefore it is not clear that those solutions could lend themselves to overcoming the uplink
bottleneck.  

A more related solution to our work is ``link-alike''~\cite{link-alike}, which
also suggests uplink aggregation via multiple APs, but it is designed specifically for UDP
based large file transfer only. We have built a ``link-alike'' based prototype and
evaluated it for TCP transfer experiments. The aggregated TCP throughput was much poorer
compared to the efficiency of UDP transmission. The degradation of TCP throughput was due
to the wireless lossy links between the client and the participating APs, and the ``uncontrolled''
order of packets forwarded by the APs. Simple aids based on coding or buffering schemes, however,
do not solve the issues (see Section~\ref{}), which are inherently rooted from the \emph{unreliable} broadcast
link between the client and the APs. Furthermore, ``link-alike'' requires heavy modifications of
not only the APs and destination, but the clients as well, therefore renders itself hard to
deploy in practical scenarios where client devices such as mobile phones are not desired 
to be ``touched''.
}

\ignore{
While contemporary broadband downlinks offer sufficient bandwidth for
most applications, the protocol presented in this paper, ``\mytit'',
specifically addresses aggregation of highly limited uplinks.
Besides, for the sake of the applicability and ease of adoption,
\mytit targets a solution \emph{transparent} to clients, requiring
only minimal modifications.  Moreover, \mytit targets a generic
support for both existing transport layer protocols, i.e., UDP
\emph{and} TCP, to support a wide range of popular applications such
as large file transfer, streaming, etc.}

\ignore{To have a better understanding of our design considerations and
motivations, we first present an overview of the BaPu architecture and
two example application scenarios.  We will also summarize the key
features of BaPu and our experimental results.}

\ignore{
%\item
 \textit{TCP friendly}: Contrary to previous work, \mytit allows
  not only UDP, but also TCP multiplexing by employing a novel
  mechanism, \emph{Proactive-ACKs}. This renders uplink aggregation
  TCP friendly. Based on this technique, \mytit supports not only
  large file upload, but also, e.g., HD video streaming.
}
\ignore{\item \textit{Support large file transfer and streaming}:
%Thanks to our client transparent design and support for both TCP and UDP, 
BaPu can work with a large set of 
applications out-of-the-box. Prior work only supports UDP based large file transfer. 
BaPu also supports HD video streaming. }

%\end{itemize}

\section{System Overview}

\subsection{Application Scenarios}

\begin{figure}
\centering
\includegraphics[width=\columnwidth]{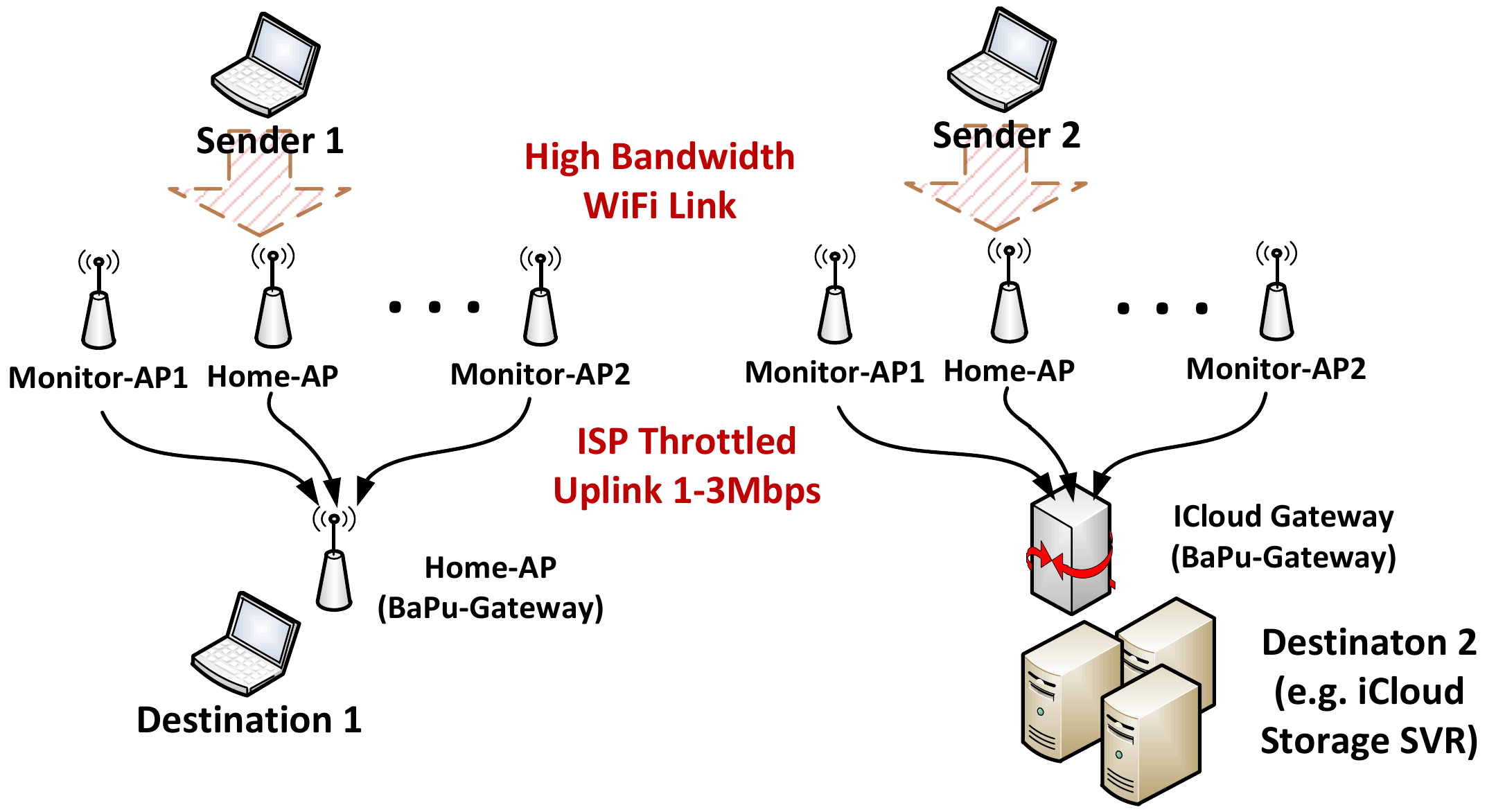}
\caption{\label{fig:bapu-arch} \mytit system architecture and two
  example application scenarios. 
Scenario 1 (left): Sender~1 shares an HD video
  with a remote end user through a \mytit-enabled Home-AP and
  neighboring Monitor-APs. % The \mytit-Gateway runs on the Home-AP of a destination user;
Scenario 2 (right): Sender 2 backs up a large file
  to iCloud through a \mytit-enabled Home-AP and Monitor-APs.
  %The  \mytit-Gateway runs on a gateway server deployed ``in front'' of the cloud storage servers;
  }

\end{figure}

For ease of understanding, we first introduce two typical application
scenarios that benefit from \mytit -- see Figure~\ref{fig:bapu-arch}.

\textbf{Scenario 1: Instant Sharing of HD Video:} In order to
retain the control of personal content, Sender~1 shares his HD
video directly from his hard drive and streams it instantly, i.e., in
real-time, to the other user -- Destination~1. Both users are
connected to their Home-APs, with an uplink connection from Sender~1
to Destination~1 throttled to 1 $\sim$ 3Mbps by Sender~1's ISP. The HD
video has 8Mbps playback rate (standard 1080p video bit rate), so
Sender~1's single uplink cannot handle this content in
real-time. However with \mytit, the idle uplink of the neighboring
Monitor-APs are exploited to boost the uplink throughput. The
\gw, the Home-AP of Destination 1, plays the role as
the \emph{aggregator} to receive and forward multiplexed traffic to
Destination~1.

\textbf{Scenario 2: Instant Backup of Large File:} Sender~2
wishes to backup his HD video clip to some cloud storage service such
as iCloud.  With the 3Mbps uplink rate, it takes over an hour to
upload a 30 minute HD video.  With \mytit, neighboring Monitor-APs and
Home-AP upload data in parallel.  iCloud just needs to deploy a
gateway server in front of the cloud storage servers. This gateway
server runs the \gw software to handle parallel uploading
from multiple paths. Using \mytit, file uploading time is greatly
reduced.

\textbf{Security and Privacy:} In both application scenarios, the APs are configured
to have two SSIDs, an encrypted SSID (i.e., WPA) for private use, and an unencrypted (open) SSID. The \mytit traffic is carried over neighbouring unencrypted SSIDs with end-to-end security (e.g., SSL/TLS), while allowing the main SSID of participating APs to remain encrypted.

\subsection{{\large\mytit} Description}

\begin{figure}
\centering
\includegraphics[width=0.8\columnwidth]{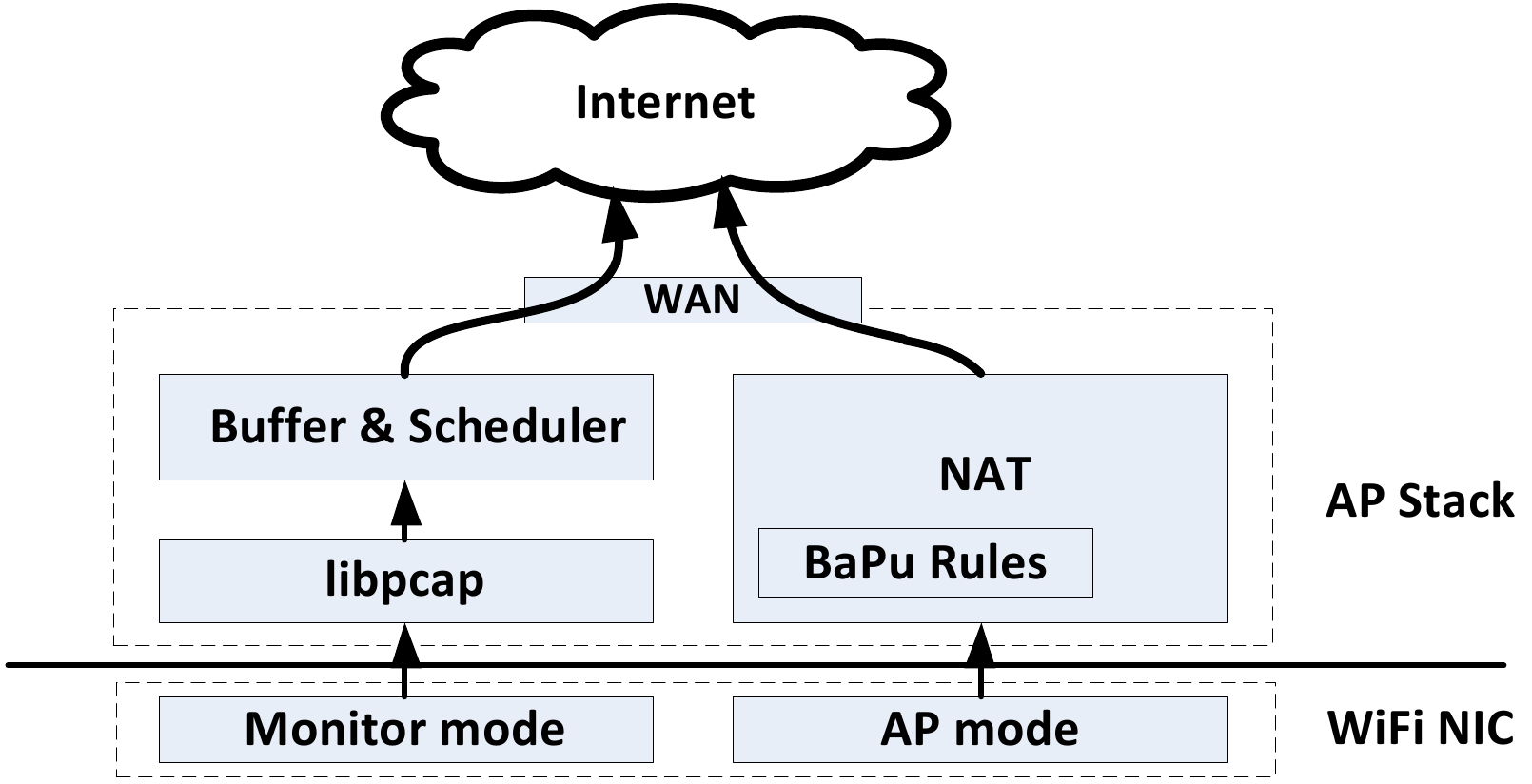}
\caption{\label{fig:bapu-ap}\mytit-AP Building components.}

\end{figure}

First, we introduce the notation used in this paper:
\noindent
\begin{itemize}
  \setlength{\itemsep}{0pt}
  \setlength{\parskip}{0pt}
  \setlength{\parsep}{0pt}
\item{\s}: device uploading data to a remote destination.
\item{\d}: device receiving uploaded data from a remote sender.
\item{\home}: AP which \s is associated to. 
\item{\mon}: in physical proximity to \d
  and \home, a couple of neighboring APs, the \mon{}s, run in
  \emph{monitor mode} on the same WiFi channel as \home.
\item{\gw}: gateway device connected to
  \d. As shown in Figure~\ref{fig:bapu-arch}, \gw can
  be the WiFi AP of the \d or the gateway server at the edge of
  some cloud data center.
\item{\ap}: for abstraction, \home and
  \mon will be called \ap, thereby
  representing the role that APs play in a \mytit data upload session.
\end{itemize}

In \mytit, \s is associated with its \home, and the uploading of data
is aggregated via unencrypted wireless link. The data, however, are
protected with some end-to-end security mechanism (e.g., SSL/TLS).
\home{} and \mon{} are configured
to run in \emph{both} WiFi AP mode and WiFi monitor
mode\footnote{Modern WiFi drivers, such as the prominent
  {\texttt{Ath9k}} family, allow one physical WiFi interface to
  support running in multiple modes}. The general \ap{} setup is
illustrated in Figure~\ref{fig:bapu-ap}.

The WiFi link between a \s{} and its \home{} generally provides high
bandwidth, up to hundreds of Mbps with 802.11n. The link between a
\ap{} and the \d{}, however, is throttled by the ISP.  At the remote
end, we place a \gw{} immediately before the \d{}.  The connection
between the \gw{} and the \d is a wired or wireless
high-speed link.  Note that being in physical proximity, unicasts
between \s{} and \home{} (AP mode) can be overheard by (some of) the
neighboring \mon{}s (monitor mode).

At a high level, \mytit is a centralized system with the controller
residing at \gw.  \mytit provides an aggregation of uplink bandwidth
that is carried out as follows.

\begin{figure}
\centering
\includegraphics[width=0.9\columnwidth]{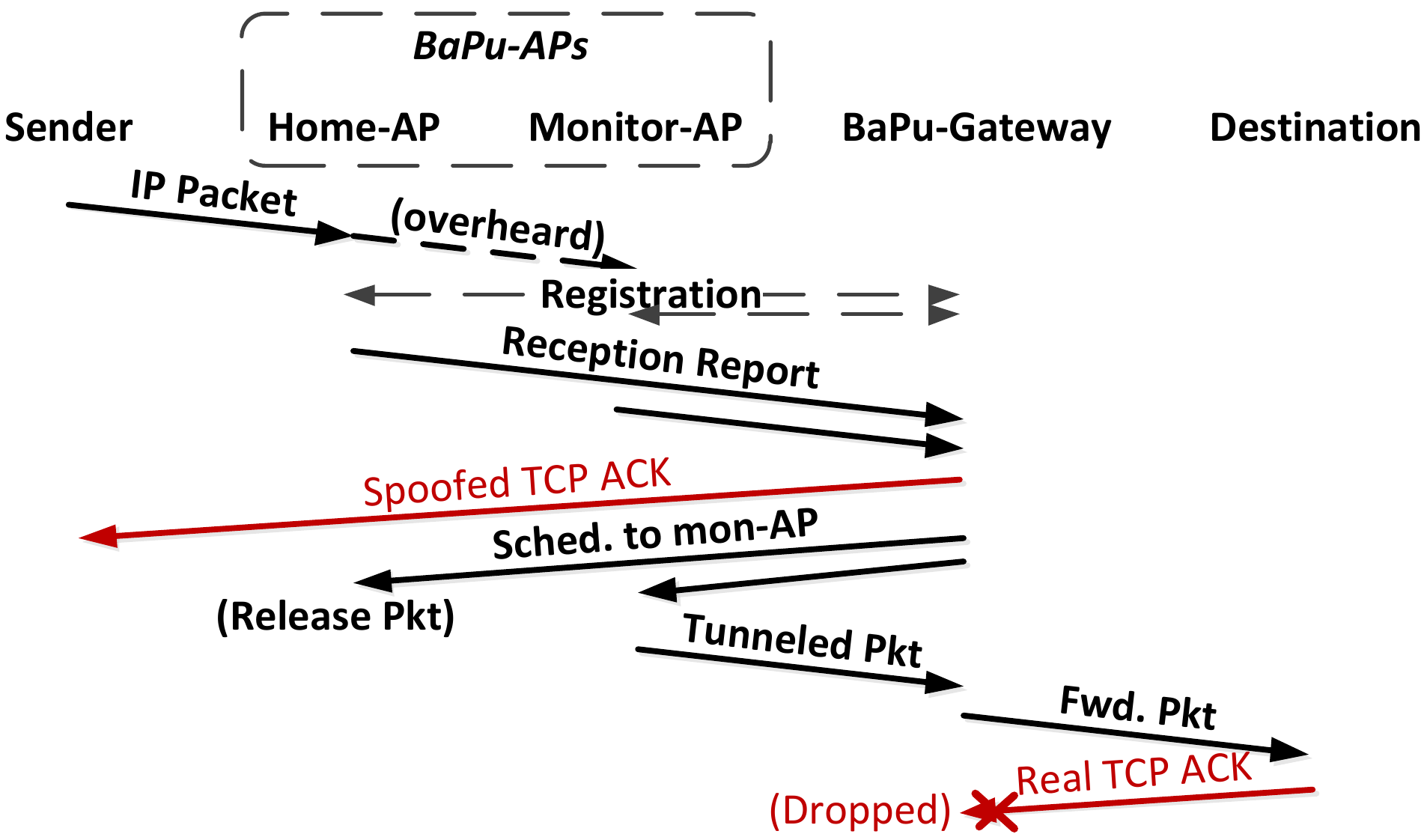}
\vspace{-0.1in}
\caption{\mytit Protocol Traffic Flow. The ACKs (red color) are managed for TCP only.}
\label{fig:bapu-pro-flow}
\end{figure}

\begin{enumerate}
  \setlength{\itemsep}{0pt}
  \setlength{\parskip}{0pt}
  \setlength{\parsep}{0pt}

\item \label{step:init}\s{} starts a TCP or UDP upload session to \d{} through its
  \home{} via WiFi.

\item \label{step:identify}\home{} and \mon{} overhear WiFi packets and identify if this
  can be a ``\mytit" session by checking the destination IP and
  port. In our prototype described in Section~\ref{sec:evaluation}, we choose a
  specific UDP/TCP port for all traffic that allows bandwidth
  aggregation.

\item \label{step:register}\ap{}s register themselves as a \emph{contributor} to \gw.

\item In \mytit, \home{} and \mon{} collaborate to upload data for
  \s{}, following a schedule that is determined by \gw{}. We will
  explain this scheduling mechanism and protocol details later in
  Section~\ref{sec:bapu}.
\ignore{Once \home{} identifies a \mytit
    session for \s{}, it dynamically updates its {\tt iptables} rules to
    prevent the \mytit session from being forwarded to WAN along the
    default route.}

  Practically speaking, \home{} and \mon{} now capture \s{}'s packets
  with {\tt libpcap} from the monitor mode, and stores them in a
  buffer.

\item \label{step:overhear}For each packet overheard, \home{} and \mon{} send packet
  reception reports to the \gw{}.

\item \label{step:report-udp}For an \emph{UDP} session, on reception of the reports, \gw{}
  determines which \ap{} will forward the captured packet in step~\ref{step:forward}.
  A scheduling message is then prepared to include
  the selected \ap{}'s identity, and this scheduling message is
  \emph{broadcast} back to all \ap{}s participating in the current
  session.

\item \label{step:report-tcp}A \emph{TCP} session is much more challenging to support than
  UDP. To properly multiplex \s{}'s single TCP flow through multiple
  paths, \gw{} adopts a mechanism we call \emph{Proactive-ACK} mechanism. \gw{}
  sends spoofed TCP ACKs to \s{} as the \mytit session goes
  on. \emph{Proactive-ACK} is designed to make \mytit work
  efficiently with legacy TCP congestion control. We will explain
  details later in Section~\ref{sec:bapu}.

\item \label{step:forward}The scheduled AP forwards the buffered packets to \gw{}, which
  forwards to \d{}.

\item \label{step:downlink}Any downlink traffic from \d{} to \s{} just follows the default
  network path, i.e., from \d over \gw and \home to \s. 
\end{enumerate}

Figure~\ref{fig:bapu-pro-flow} shows \mytit's protocol flow.  In the
following section, we now discuss \mytit's research challenges in
details and give an insight to our design decision at each step of the
protocol.

\section{Bapu}
\label{sec:bapu}
In this section, we describe the whole \mytit system in details.
We discuss technical challenges arising and propose solutions to
achieve an efficient and practical aggregation system. We remark that \mytit shares some similarities
in the high-level architecture with prior work (e.g., Link-alike~\cite{link-alike}, FatVAP~\cite{KandulaLBK2008}),
which presented neat systems for aggregating the bandwidth among APs. However, from the practicality
aspects, the applicability of those systems is still limited due to constraints such as heavy modification of
client devices or support for only specific applications (e.g., large file transfer).
Yet our ultimate goals of the \emph{transparency} for the users, and the \emph{high-throughput} transmission for all kinds of user applications
require a new solution with unique characteristics.

\subsection{Network Unicast}
First, the transparency goal requires that legacy transport protocols be usable for
data transmission from \s to \d. Accordingly, the \s must be able to transmit data to the \d
via \emph{network unicast} through its \home.
The second reason for the need of network unicast is to increase the reliability of the transmission,
because \mytit supports TCP, whose performance depends on
the reliability of the underlying MAC layer. To be clearer, according to the IEEE 802.11 standard,
a packet with a broadcast destination address is transmitted only once by the WiFi
device, whereas up to 12 MAC-layer retransmissions are tried for a failed unicast destination address,
therefore a unicast is much more reliable than a broadcast.
Consequently, supporting network unicast is an essential requirement in \mytit, while in prior work~\cite{link-alike},
broadcast is preferred due to the simplicity goal of their system.

\vskip 1eX\noindent{\bf Packet Overhearing:}
In WiFi network, although both network unicast and network broadcast use the same method of wireless broadcast
to transfer data, the difference lies in the MAC layer, where the next-hop physical
address is specified to the unicast address or broadcast address.
This complicates the packet overhearing capability at \mon{}s.
As \home is the first hop in the transmission path, the \s, according to the
underlying routing protocol, has to use the \home's physical address
as the next-hop address in the 802.11 header. While \home as a next hop can receive the packet,
\mon{}s automatically discard the packet due to mismatched physical address.
Therefore, barely relying on the default network behavior does not let \mon{}s capture
packets sent by \s{}s in other WLANs.

\mytit's solution is to configure \ap{}s to operate simultaneously
in two modes: \emph{AP mode} and \emph{monitor mode}. The former mode is used for
serving clients in the AP's own WLAN, whereas the latter is used for overhearing packets
in the air. In monitor mode, packets are
captured in raw format via the use of {\tt libpcap}.

\vskip 1eX\noindent{\bf Packet Identification:}
Each packet sent from the \s (step~\ref{step:init}) contains the session information in the packet's IP header such as
the protocol identifier, the source and destination IP addresses and ports. With this information,
\home can uniquely identify the \s (step~\ref{step:identify}). In contrast, \mon{}s may have ambiguity in identifying the \s,
as \s{}s from different WLANs may (legally) use the same IP address.
To resolve such conflict, we write a frame parser for the packet's MAC header
to obtain the {\tt BSSID} that identifies the WLAN
the session belongs to. Therefore, any session in \mytit is now uniquely determined on the following 6-tuple
{\tt <BSSID, proto, srcIP, dstIP, srcPort, dstPort>}.

\vskip 1eX\noindent{\bf Duplicate Elimination:}
As mentioned earlier, unicasting a packet may involve a number of (MAC-layer) retransmissions due to
wireless loss occurred between the \s and its \home. This benefits the data transmission between them.
Nevertheless, it is possible that a nearby \mon can overhear more than one (re)transmission
of the same packet, which creates duplicates and floods the \mon's uplink if all the retransmitted packets
get scheduled. To identify the duplicate packets, we keep records of {\tt IPID} field in the
IP header of each overheard packet. Since {\tt IPID} remains the same value for each MAC-layer retransmission,
it allows \mon{}s to identify and discard the same packet. It is worth noting that in TCP transmission,
the TCP sequence number is not a good indicator to identify the duplicate packets, as it is unique for
TCP-layer retransmitted packets, but not unique for MAC-layer retransmissions.

\subsection{Tunnel Forwarding}
The transparency goals requires that the \s's data transfer session is unaware of
the aggregation protocol in \mytit. A seemingly straightforward solution
is that \home and \mon{}s forward the \s's packets with spoofed IP addresses.
It is, however, impractical for two reasons:
1) many ISPs block spoofed IP packets;
2) forwarded packets by \mon{}s are unreliable, because they are raw packets
overheard from the air. Our approach is that each \ap 
conveys the \s's data via a separate TCP tunnel.
Since we support a transparency for aggregation over multiple paths,
the techniques for tunnelling and address resolving in each single path require
a careful design at both \ap{}s and \gw.

\vskip 1eX\noindent{\bf Tunnel Connection:}
Once a \ap identifies a new \s-\d session (step~\ref{step:register}) based on the 6-tuple, it establishes a tunnel connection
to \gw. Regardless of the session protocol, a tunnel connection between the
\ap and \gw is always a TCP connection. The choice of TCP tunnel is partially
motivated by the \emph{TCP-friendliness}. We desire to aggregate the idle bandwidth
of \ap{}s without overloading the ISP networks. Besides, since TCP tunnel can provide a reliable
channel, it helps keep a simple logic for handling a reliable aggregated transmission.

\vskip 1eX\noindent{\bf Forwarding:}
In the registration (step~\ref{step:register}) to \gw, the \ap receives an {\tt APID} as its ``contributor'' identifier for
the new session. The {\tt APID} is used in all messages in the protocol. 
Both control messages (registration, report, scheduling) and data messages are exchanged via
the TCP tunnel, which ensures reliable transmissions.
On reception of a scheduling message with matching {\tt APID}, the \mon encapsulates
the corresponding \s's packet in a \mytit data message and sends it to \gw (step~\ref{step:forward}), which then
extracts the original data packet, delivers to the \d.
\ignore{At the same time, \gw broadcasts back to all \ap{}s
a \mytit acknowledgement for reception of the data. The \mon{}s, who are not
the selected forwarder for the corresponding packet, keep the captured packet
in their buffer until a timeout or a \mytit acknowledgement is received.
This helps the \gw to schedule to another \ap in case the selected \ap is suddenly offline.
In TCP session, if no other \ap{}s have the corresponding data packet, the \d
will automatically trigger the TCP retransmission algorithm. The TCP semantics will
be discussed in more details in Section~\ref{sec:proack}.}
In \mytit, the control messages are short, thus introducing only a small overhead in the backhaul.
%This is important to the efficiency of the system, which we will evaluate in Section~\ref{sec:evaluation}.

\vskip 1eX\noindent{\bf NAT:}
In WiFi network, the \s is behind the \home and the \s might also reside behind
a gateway. By default, a NAT (network address translation) is performed
for the session between the \s and the \d.
In \mytit, the \s's data are conveyed to the \d via separate tunnels from
each participating \ap. Therefore, different from the address translation
in a traditional network, \mytit requires that the NAT mapping information of the end-to-end
session must be known
to transfer the embedded data to the desired \d. Consequently, in the registration step, each \ap,
besides {\tt APID}, also receives the NAT mapping records from \gw.
%Please note that the NAT operation is not mandatory at AP side. We can shift the NAT operation to the BaPu-Gateway.  

Besides, since the downlink capacity is enormous, we allow all reverse (downlink) traffic from \d to \s to traverse along the \emph{default downlink path}. In addition, as there might be multiple tiers of NAT boxes in the middle, we must ensure that the NAT mapping for a session is properly installed on all NAT boxes along the path between \s and \d in order for the returning traffic to traverse the NAT boxes properly.  Therefore, the first few packets in a session must go along the \emph{default uplink path}.  This means the first packet in UDP sessions or the 3-way handshake traffic in TCP sessions are not tunnelled.

%\fixme{how about iptables, conntrack?}

\subsection{TCP with Proactive-ACK}
\label{sec:proack}
TCP ensures successful and in-order data delivery between the 
\s and the \d. 
\ignore{TCP relies on two major mechanisms, flow control and congestion control. 
The former one prevents the sender from overrunning the receiver buffer. The latter one prevents the sender from overrunning the network path between the sender and the receiver.}
 In TCP, each packet is 
identified with a sequence number and must be acknowledged by the \d
to indicate the proper delivery. The \s maintains a dynamic CWND (congestion window) 
during the on-going session, which indicates the maximum number of packets that can be sent on 
the fly, therefore determines the TCP throughput.

The \s's CWND size is affected by the acknowledgements received from the \d.
First, the growth rate of CWND depends on the rate of receiving acknowledgements, i.e., the link latency.
Second, missing acknowledgement within a RTO (retransmission timeout) causes
the \s to issue a \emph{retransmission}. On the other side, if the \d receives some out-of-order sequence,
%this implies that the packet corresponding to the missing sequence number is lost or delayed due to congested network. As a result, the \d
it sends a DUPACK (duplicate acknowledgement) 
to inform the \s of the missing packet. By default~\cite{Allman:2009:TCC:RFC5681}, the \s will issue
a \emph{fast retransmission} upon receiving 3 consecutive DUPACKs.
Both retransmission and fast retransmission cause the \s to cut off
the CWND accordingly to slow down the sending rate and adapt to the congested 
network or slow receiver.

\vskip 1eX\noindent{\bf Performance issues with aggregation:}
TCP was designed based on the fact that the out-of-order 
sequence is generally a good indicator of lost packets or congested network.
However, such assumption no longer holds 
in \mytit.
\begin{itemize}
  \setlength{\itemsep}{0pt}
  \setlength{\parskip}{0pt}
  \setlength{\parsep}{0pt}
\item \emph{Out-or-order packets:} In \mytit, the packets belonging to the same TCP session are \emph{intentionally} routed through 
multiple \ap{}s via diverse backhaul connections in terms of capacity, 
latency, traffic load, etc. This results in \emph{serious} out-of-order sequence at 
\gw, which eventually injects the out-of-order packets to the \d.
\item \emph{Double RTT:}
Also, due to the aggregation protocol, data packets in \mytit are delivered to the \d with a double round-trip-time (RTT)
compared to a regular link. This causes the \s's CWND to grow more slowly and peak at lower values.
\end{itemize}
Consequently, with an \emph{unplanned} aggregation method, the TCP congestion control mechanism is \emph{falsely} triggered,
resulting in considerably low throughput. As we show later in Section~\ref{sec:evaluation}, a simplified prototype of
our system, which share similarities with the system in~\cite{link-alike}, gives poor TCP throughput.

\vskip 1eX\noindent{\bf Solution:}
To address the performance issue, we first investigated a simple
approach:
data packets forwarded by \ap{}s are buffered at \gw until
a continuous sequence is received before injecting to the \d.
This solution, however, encounters the following issues:
1) \emph{Efficiency:} Introducing a buffer for each session at \gw
is wasteful of memory, since the \d already maintains
a TCP buffer for its own session. Furthermore, this does not scale well
when more simultaneous sessions are established.
2) \emph{Optimality:} Due to the difference in capacity, latency, loss rate
among backhaul uplinks, it is not clear how to determine the optimal buffer size.
3) \emph{Performance:} In fact, we implemented a buffering mechanism at \gw, and
the results (Section~\ref{sec:eval-buffer}) showed that using buffering mechanism
\emph{does not} help improving the TCP throughput.

Now we introduce a novel mechanism called \emph{Proactive-ACK}, which is
used in step~\ref{step:report-tcp} of \mytit protocol.
The principle of Proactive-ACK mechanism is to actively control
the exchange of acknowledgements instead of relying on the default
behaviour of the end-to-end session. With Proactive-ACK, we solve
both \emph{out-of-order packet} and \emph{double RTT} issues.
In the following paragraphs, we call acknowledgements
actively sent by \gw \emph{spoofed} acknowledgements, while the ones
sent by the \d are \emph{real} acknowledgements.

\vskip 1eX\noindent{\bf Managing DUPACK:}
In \mytit, most of out-of-order packets are caused by the aggregation mechanism
via multiple \ap{}s. To avoid the cutting off of the CWND at the \s, we intentionally
discard all DUPACKs received from the \d, as we observed that most of DUPACKs
generated by the \d in \mytit are due to the multiple-path aggregation.

However, by dropping DUPACKs from the \d, we need to handle the case of actual lost packets in the air
between the \s and \ap{}s.
Concretely, if the report for expected TCP sequence number is not received within certain time window,
it is implied that this sequence is lost on all participating \ap{}s.
Now that \gw sends a spoofed DUPACK back to
the \s in order to mimic the TCP fast retransmission mechanism for fast recovery from packet loss.

\vskip 1eX\noindent{\bf Managing ACK:}
Besides the effect of DUPACKs, the CWND size of the \s is also highly affected by
the double RTT introduced by the protocol. Not only the CWND grows slowly, but
the chance of CWND being cut off is also higher.
With Proactive-ACK mechanism, in step~\ref{step:report-tcp}, \gw sends back to the \s the spoofed ACK after
receiving the report from \ap{}s.
The intuition is that all the packets that are reported by some \ap{}s are currently stored in
those \ap{}s' buffer. Due to the reliability of the TCP tunnel between \ap{}s and \gw, the
reported packets will be eventually forwarded to \gw in reliable manner.  Therefore, as long as \gw
identifies a continuous range of reported TCP sequence, immediately sending a spoofed ACK back to
the \s helps maintaining a high and constant throughput, as the RTT with respect to the \s is reduced
to a value around the real RTT.
This approach prevents the cutting off of CWND at the \s.

Since \gw manually sends spoofed ACKs to the \s, on reception of real ACKs the \d,
\gw simply discards the real ACKs.

\vskip 1eX\noindent{\bf TCP semantics:}
We have two important remarks on the TCP semantics:
\begin{itemize}
  \setlength{\itemsep}{0pt}
  \setlength{\parskip}{0pt}
  \setlength{\parsep}{0pt}
\item Immediately sending the spoofed ACKs after receiving the reports
may result in spoofed ACKS being received at the \s before data packets being
forwarded to the \d. This increases the CWND
in a more aggresive manner than the standard mechanism.
\item Dropping real DUPACKs and sending spoofed DUPACKS can increase the time
for recovery of an \emph{actual} loss of packet, because the loss reflected
by the \d is not immediately signaled to the \s.
For example, if an AP who has been scheduled to forward the selected packet
is suddenly offline, it takes a longer time for the packet to be scheduled again after a timeout
and later forwarded to the \d.
\end{itemize}
Despite the slightly difference in TCP semantics, the Proactive-ACK mechanism
has been proved to give a significant improvement to the TCP throughput.
We present these results in Section~\ref{sec:evaluation}.

%%% this is for evaluation section, but put here

\begin{figure}
\centering
\includegraphics[width=\columnwidth]{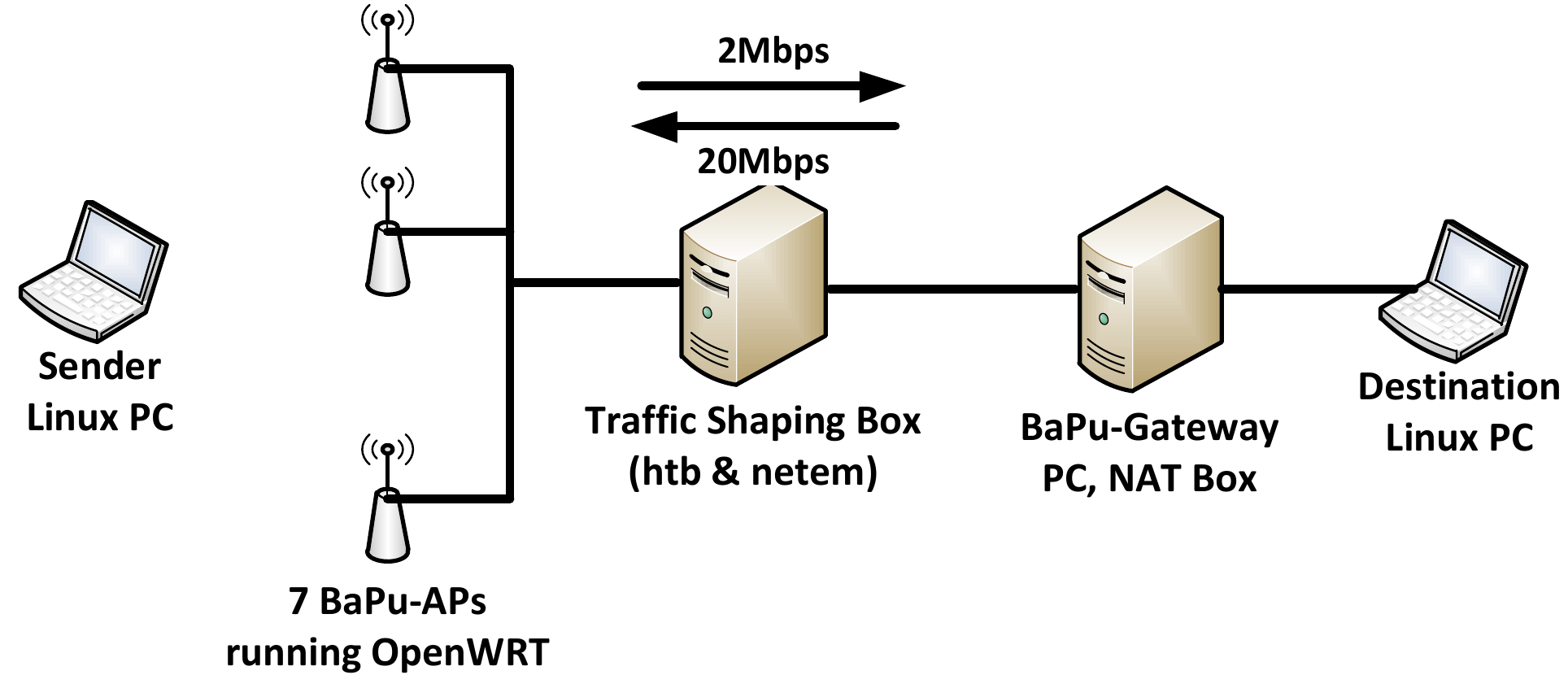}
\caption{\mytit Experiment Setup. 7 \mytit-APs and 1 \mytit-Gateway are
  inter-connected.  A traffic shaping box is set up in between to emulate
  a typical residential network setting. }
\label{fig:bapu-exp-setup}
\end{figure}

\begin{figure*}
	\centering
	\subfloat[\label{fig:basic-total-32ms}UDP and TCP throughput]{
		\includegraphics[width=0.9\columnwidth]{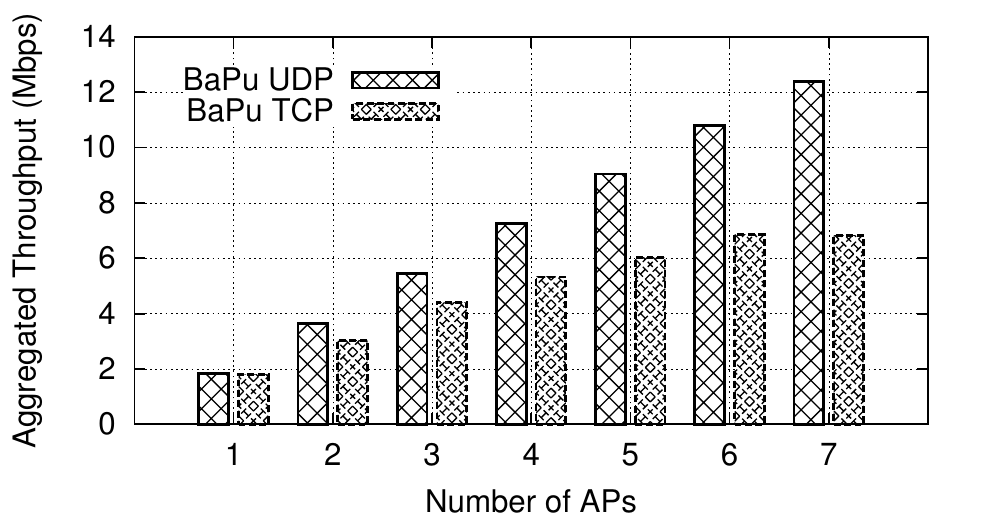}
	}
	\subfloat[\label{fig:basic-total-percent-32ms}UDP and TCP aggregation efficiency]{
		\includegraphics[width=0.9\columnwidth]{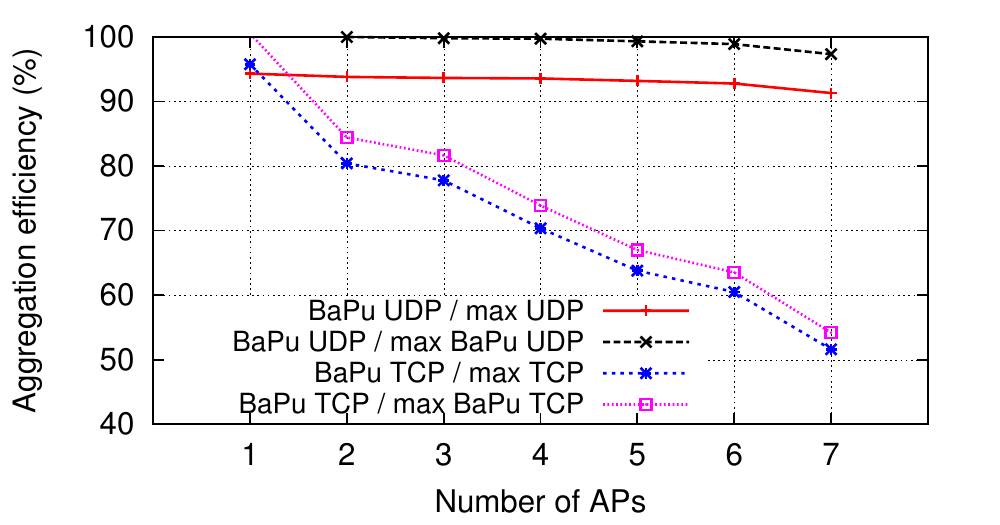}
	}
	\caption{\basic aggregation for UDP and TCP with 2Mbps 32ms RTT uplinks.}
\end{figure*}

\subsection{Scheduling}
\label{sec:schedule}
The bandwidth aggregation performance depends on the efficiency of multiplexing
data among \ap{}s to best utilize the idle uplink bandwidth.

In \mytit, we adopt a \emph{centralized scheduler} at \gw.
There are two main factors to select this design.
First, with the centralized design, it does not only simplify the implementation,
but also allow easy extension of the design with extra logic to further optimize
the scheduling strategy. Second, a scheduler usually requires complex processing
and memory capability, which might overload the \ap{}s with much lower capability
if scheduling decisions are migrated to the APs.

The scheduling strategy is based on the received reports 
in step~\ref{step:report-udp} and~\ref{step:report-tcp} of the protocol. Each report from a \ap contains
a sending buffer size obtained from the Linux kernel via {\tt ioctl()} function call.
This value specifies how much a \ap can contribute to the aggregation.
Based on reports, \gw applies First-Come-First-Served strategy to select
a forwarder among \ap{}s who have captured the same packet. This approach is
similar to those applied in~\cite{link-alike,KandulaLBK2008}.
The rationale for choosing this approach are
\begin{itemize}
  \setlength{\itemsep}{0pt}
  \setlength{\parskip}{0pt}
  \setlength{\parsep}{0pt}
\item \emph{Fairness:}
Sharing bandwidth for aggregation takes into account the available bandwidth
of participating \ap{}s, as because AP owners have
different subscription plans.
\item \emph{Protocol independence:} Scheduling decision is made based on
the \ap{}s' sharing capability, not on the particular transport protocol.
\end{itemize}

\ignore{
\subsection{Embedded devices}

\fixme{Challenges in programming with OpenWRT, cross-compile, reflash}

\fixme{Memory constraints lead to what difficulties?}

\subsection{Discussion}
\vskip 1eX\noindent{\bf Availability of APs:} from reviewer: "80\% of them are secure" and "operate on other channels"

\vskip 1eX\noindent{\bf Double latency issue:} latency aware applications

\vskip 1eX\noindent{\bf TCP semantics:} what if an AP who got scheduled but fails before forwarding?

\vskip 1eX\noindent{\bf Security and Privacy issues:} end-to-end encryption should be enough

\vskip 1eX\noindent{\bf Packet loss:} strawman approach: coding or modulo scheduling do not work

\vskip 1eX\noindent{\bf Out-of-order packets:} strawman approach: buffering do not work

(mention the above strawman approaches very briefly, leave the details in the evaluation section where we can show the charts)
}
\section{Evaluation}
\label{sec:evaluation}

\begin{table}
\centering
\begin{tabular}{|l|c|}
\hline
Distance & RTT\\
\hline\hline
Regional: 500 - 1,000 mi& 32ms~\cite{akamai:hd}\\
\hline
Cross-continent: $\sim$ 3,000 mi& 96ms~\cite{akamai:hd}\\
\hline
Multi-continent: $\sim$ 6,000 mi& 192ms~\cite{akamai:hd}\\
\hline
inter-AP in Boston& 20ms $\sim$ 80ms\\
\hline
\end{tabular}
\caption{\label{tab:latency} Network RTT Latency. Inter-AP RTT
  measured by our Open Infrastructure WiFi testbed in greater Boston,
  representing typical
  RTT between home APs, covering Comcast, RCN, and Verizon~\cite{open-infrastructure}.} 
\end{table}

In this section, we evaluate the performance of \mytit for UDP
and TCP in various system settings.

\vskip 1eX\noindent{\bf Experiment Setup:} Our experiment setup is
shown in Figure~\ref{fig:bapu-exp-setup}.  Our testbed consists of a
\s, 7 \ap{}s, a \gw, a \d node, and a traffic shaping box. All APs are
Buffalo WZR-HP-G300NH 802.11n wireless routers. This router model has
a 400MHz CPU with 32MB RAM.  We reflashed the APs with
OpenWRT firmware, running Linux kernel 2.6.32 and
{\texttt{ath9k}} WiFi driver. In our experiments, we select one \ap to
act as a \home which the \s is always associated to. The other
6 \ap{}s act as \mon{}s to capture the traffic in monitor mode. The
\gw runs on a Linux PC, and the \d runs behind the \gw. 
The \s and the \d are both laptops with 802.11n WiFi card,
running the standard Linux TCP/IP stack.

To emulate traffic shaping as with residential broadband, we use the 
traffic shaping box between the \ap{}s and \gw. We use Linux' {\tt
  iptables} and {\tt tc} with the {\tt htb} module to shape the
downlink bandwidth to 20Mbps and the uplink to 2Mbps.  Also,
to emulate network latency between \ap{}s and \gw, we use {\tt
  netem} to shape the RTT with different values.  The bandwidth
and latency parameter are selected to represent the typical bandwidth
capacity and regional latency in residential cable broadband that we
have measured in Boston's urban area (Table~\ref{tab:latency}).

\begin{table}
\centering
\begin{tabular}{|l|c|}
%\hline
%Uplink Capacity & 2 Mbps\\
%\hline \hline
\hline
max UDP & 1.94 Mbps\\
\hline
max \mytit UDP & 1.82 Mbps\\
\hline
max TCP & 1.9 Mbps\\
\hline
max \mytit TCP & 1.8 Mbps\\
\hline
\end{tabular}
\caption{Maximum theoretical goodput for UDP and TCP with and without \mytit overhead. Data payload size is 1350Bytes. Uplink capacity is 2Mbps.}
\label{tab:limit}
\end{table}

In our experiments, we issue long-lived 30 minutes {\tt iperf} flows
(both TCP and UDP) from \s to \d.  We choose 1350Byte as TCP/UDP
payload size in our {\tt iperf} test to make sure that the whole
client IP packet can be encapsulated in one IP packet while an \ap{}
sends it through its TCP tunnel.  All throughput values reported in
our experiment are the {\tt iperf} throughput, which is the
\emph{goodput}.

In the evaluation, we compare throughput of UDP and TCP in different
system scenarios. More precisely, we evaluate the following scenarios:
\begin{itemize}
  \setlength{\itemsep}{0pt}
  \setlength{\parskip}{0pt}
  \setlength{\parsep}{0pt}
\item \basic: \mytit system without any buffering or Proactive-ACK
  mechanism.
\item \buffering: \mytit system without Proactive-ACK mechanism, but
  enhanced by buffering at \gw.
\item \proack: this is the full \mytit system.
\end{itemize}

\ignore{
In the above scenarios, in order to have a practical testbed environment,
we keep other features in \mytit such as network unicast, tunnel forwarding
with NAT-ing, centralized scheduler at \gw.}

\begin{figure}
\centering
\includegraphics[width=\columnwidth]{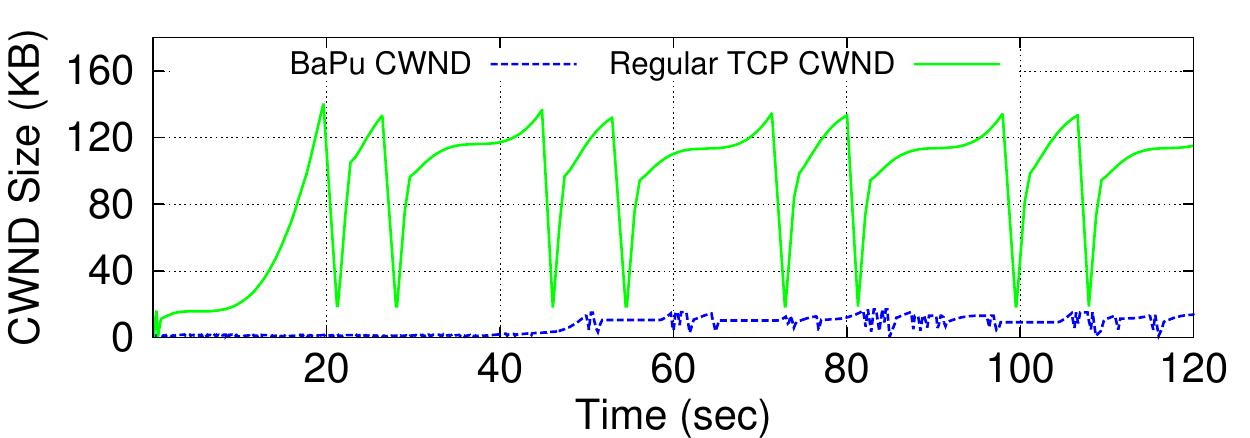}
\caption{Sender's TCP CWND growth compared between \basic and regular
  single AP with 32ms RTT and total 14Mbps uplink.}
\label{fig:plot-basic-7ap-32msrtt-5min-iperf-tcpinfo-cwnd-retrans}
\end{figure}

\subsection{{\Large{\basic}}: Efficient UDP, Poor TCP}
\label{sec:eval-basic}
\ignore{
We evaluate the performance of \basic by deploying the testbed
with number of \ap{}s increasing from 1 to 7.}

\subsubsection{System efficiency with UDP throughput}
The practicality of \mytit lies in its efficiency.  In contrast to
related work, \mytit's transparency goal, not requiring any
modifications at the client side, has motivated the design of \mytit's
underlying technical details. We now first measure \mytit's efficiency
by the throughput with UDP, as it provides a light-weight end-to-end
transmission between \s and \d.  Figure~\ref{fig:basic-total-32ms}
shows the achieved aggregated UDP throughput with numbers of
participating \ap{}s increasing from 1 to 7. We observe that the aggregated UDP throughput
increases proportionally with the number of \ap{}s, and achieves 12.4Mbps with 7 \ap{}s. 
To put this figure into
perspective, note that related work by~\citet{link-alike} achieves
similar UDP throughput but without support for TCP or client transparency.

\subsubsection{Low TCP throughput}
We conduct the same experiments also for TCP
transmission. Figure~\ref{fig:basic-total-32ms} shows that the
aggregated TCP throughput does not benefit much when the number of
\ap{}s increases. The TCP aggregated throughput is always lower than
the UDP's in the same setup, and the gap between UDP and TCP
performance increases along with the number of \ap{}s. For example, we
achieve only 6.83Mbps with 7~\ap{}s.

\subsubsection{Aggregation efficiency}
In addition to measuring aggregated throughput, we evaluate our system
based on another metric, \emph{aggregation efficiency}.  We define
\emph{aggregation efficiency} as the ratio between practical
throughput over the maximum theoretical goodput.  Due to the TCP/IP
header and \mytit protocol overhead, the actual goodput is less than
the uplink capacity.  With all protocol header overhead accounted, we
derive the maximum theoretical goodput as the given backhaul capacity
of 2Mbps.  Table~\ref{tab:limit} lists the maximum throughput when
data is transmitted via standard UDP/TCP and via \mytit.

As shown in Figure~\ref{fig:basic-total-percent-32ms}, \basic UDP can
harness close to 100\% idle bandwidth. Even if we consider the extra
overhead incurred by \mytit protocol messages, UDP aggregation
efficiency is still over 90\% in all cases.  In contrast, the
aggregation efficiency for TCP degrades quickly as more \ap{}s join
the cooperation. With 7 \ap{}s, \basic transforms only 50\% of idle
bandwidth to effective throughput.

\begin{figure}
\centering
\includegraphics[width=0.9\columnwidth]{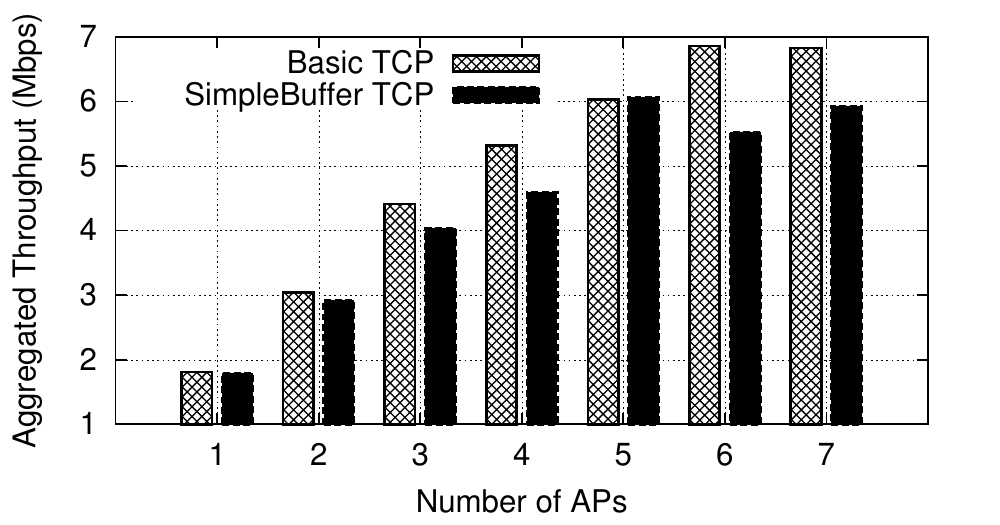}
\caption{\basic vs. \buffering comparison in TCP throughput with 2Mbps
  32ms RTT uplinks.}
\label{fig:plot32ms-compare-basic-buffering}
\end{figure}

\begin{figure*}
\centering
\subfloat[\label{fig:proack_tcp_throughput} 
Aggregated TCP throughput]{\includegraphics[width=0.9\columnwidth]{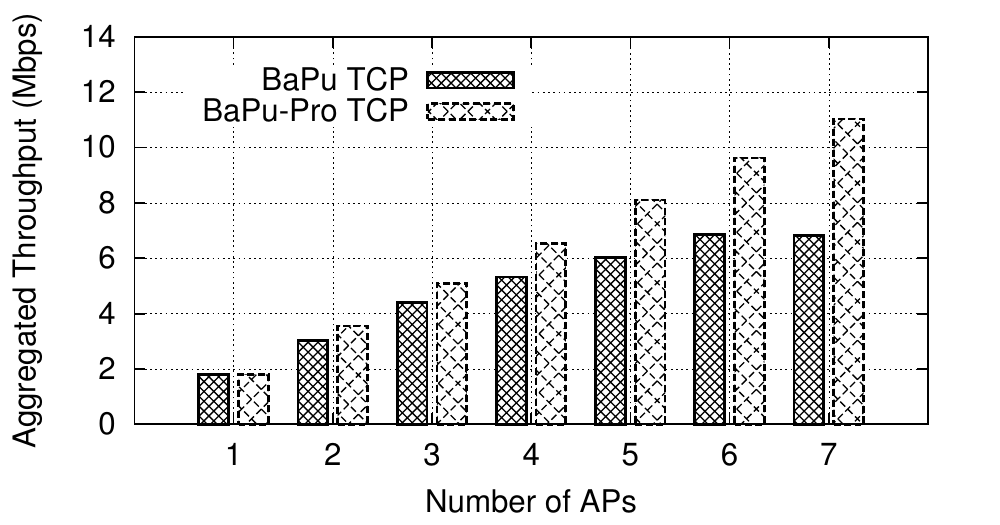}\vspace{-0.5cm}}
\subfloat[\label{fig:proack_tcp_th_vs_upper_limit} Aggregation efficiency]{\includegraphics[width=0.9\columnwidth]{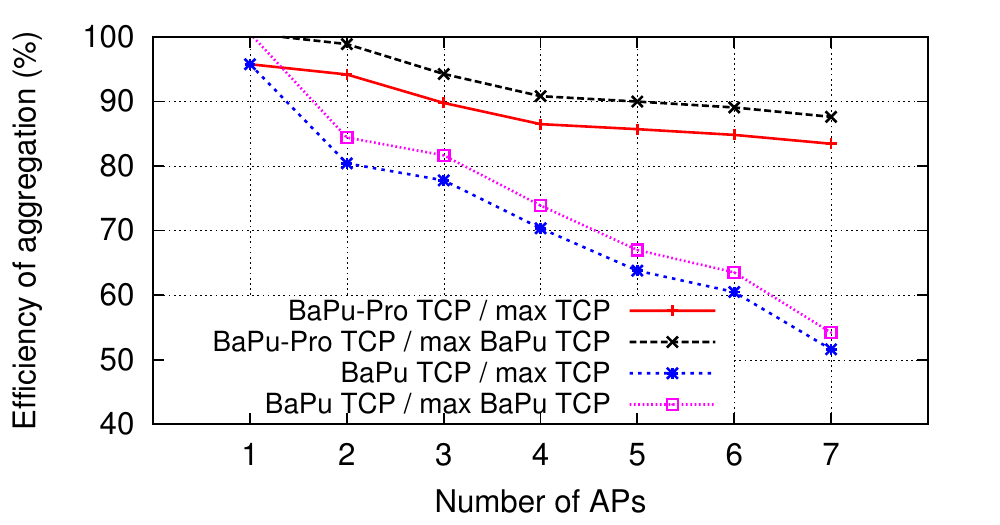}\vspace{-0.5cm}}
\vspace{-0.2cm}
\caption{\proack vs. \basic: comparison with 2Mbps 32ms RTT uplinks.}
\end{figure*}

\begin{figure*}
\centering
\includegraphics[width=1.9\columnwidth]{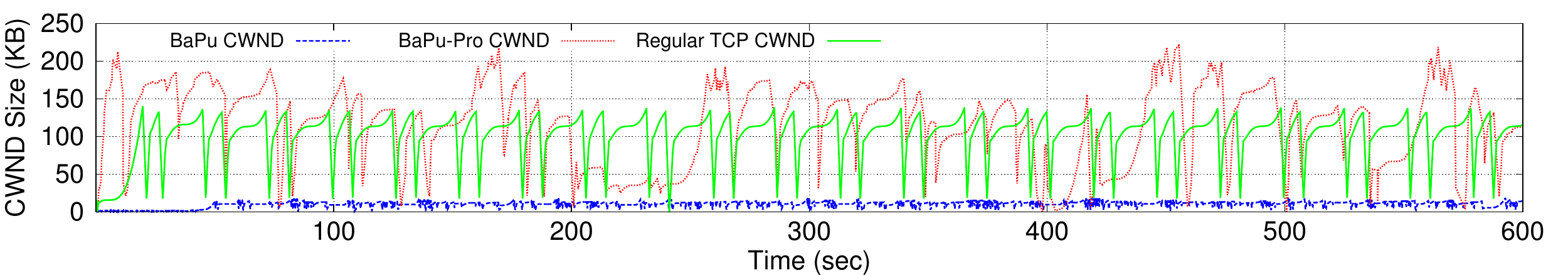}
\caption{TCP sender CWND growth comparison: \proack vs. \basic vs. normal TCP.}
\label{fig:tcp_info_basic_vs_proack}
\end{figure*}

\subsubsection{Discussion: \basic's poor TCP performance}
\label{sec:eval-basic-cwnd}
We can observe several factors in Section~\ref{sec:proack} that
decrease the aggregated TCP throughput. In this section, we carry out
an analysis on the \s's CWND size in \basic.  To justify our analysis,
we inspect the TCP behavior by examining the Linux kernel TCP stack
variables. We call {\tt getsockopt()} to query the {\tt TCP\_INFO}
Linux kernel data structure.  {\tt TCP\_INFO} includes the system time
stamp, the \s's CWND, number of retransmissions, etc.  We have
modified the {\tt iperf} code to log {\tt TCP\_INFO} each time {\tt
  iperf} calls {\tt send()} to write the application data to the TCP
socket buffer.

Figure~\ref{fig:plot-basic-7ap-32msrtt-5min-iperf-tcpinfo-cwnd-retrans}
shows the CWND growth in a 120 second {\tt iperf} test with 7~\ap{}s.
The theoretical throughput here $2\mbox{Mbps}\times7 = 14\mbox{Mbps}$.
In comparison, we carry out another {\tt iperf} test with standard TCP
through a single, regular AP with 14Mbps uplink capacity.  The CWND
growth in a normal TCP connection is also shown in
Figure~\ref{fig:plot-basic-7ap-32msrtt-5min-iperf-tcpinfo-cwnd-retrans}.
As shown, the \s's CWND remains at a very low level.  Our captured
packet trace at the \s shows that lots of DUPACK packets and RTO incur
a lot of retransmissions, which results in very low TCP throughput.

\subsection{Does {\Large\buffering} help?}
\label{sec:eval-buffer}
As discussed in Section~\ref{sec:proack}, a simple \emph{buffering}
mechanism does \emph{not} solve the TCP performance issue due to
difference in \ap{} uplink characteristics (latency, packet loss). In
this section, we show experimentally that a buffering mechanism cannot
help in improving the TCP throughput.  The experiment is performed for
equal uplink capacity and latency, i.e., we eliminate external factors
such as asymmetric links among \ap{}s.

Figure~\ref{fig:plot32ms-compare-basic-buffering} depicts the
throughput comparison between \basic and \buffering.  Surprisingly,
the throughput is \emph{worse} with \buffering. We have also adjusted
the buffer size at \gw, but the throughput still remains as low as
shown in Figure~\ref{fig:plot32ms-compare-basic-buffering}.  We have
investigated \s's CWND, and we have seen that it peaks at low values,
similarly to the behavior in \basic. The packet trace also shows a
lot of retransmissions.

\subsection{{\Large\proack} Performance}
\label{sec:eval-proack}
Now, we conduct a comprehensive set of experiments to evaluate the
performance of \proack. First, we validate our Proactive-ACK mechanism
by comparing \proack against \basic. Second, we measure the
performance of \proack under a variety of network settings (network
latency, wireless link quality, etc.). Finally, we demonstrate that
\proack is feasible for both, streaming and large file transfer
applications.

\subsubsection{TCP Throughput -- \proack vs. \basic}
We carry out the same {\tt iperf} test as described in
Section~\ref{sec:eval-basic} with \proack.  As shown in
Figure~\ref{fig:proack_tcp_throughput}, the aggregated TCP throughput
of \proack significantly outperforms the one of \basic.  With
7~\ap{}s, \proack achieves 11.04Mbps, i.e., 62\% improvement over
\basic. Furthermore, Figure~\ref{fig:proack_tcp_th_vs_upper_limit}
shows that \proack achieves at least 88\% aggregation efficiency in
our setup, and it achieves at least 83\% of the upper limit of
standard TCP throughput.  These results demonstrate that \proack can
achieve high aggregated throughput with high aggregation efficiency
for TCP in practical settings.

\begin{figure*}
\centering
\subfloat[\label{fig:latency-pro} Different RTT]{\includegraphics[width=0.9\columnwidth]{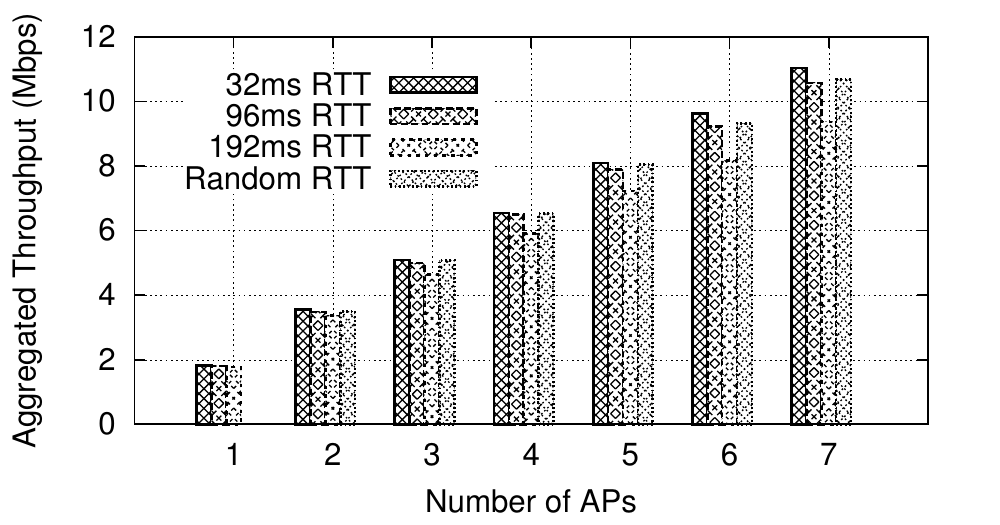}}
\subfloat[\label{fig:diversity-pro} Different packet loss rate $P$ on \mon{}s]{\includegraphics[width=0.9\columnwidth]{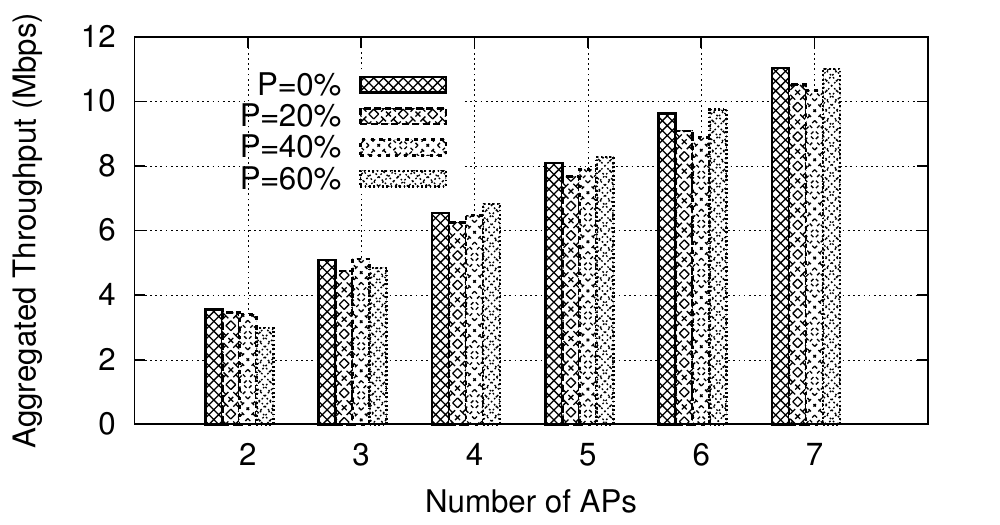}}
\caption{\proack TCP throughput.}
\end{figure*}

\subsubsection{Proactive-ACK benefit}
To justify our Proactive-ACK mechanism, we adopt the same method as in
Section~\ref{sec:eval-basic-cwnd} to examine the TCP CWND
growth. Figure~\ref{fig:tcp_info_basic_vs_proack} shows that \proack
allows the CWND to grow to very high values, contributing to the high
throughput.  For convenience, we also run a regular TCP session with a
throttled bandwidth 11Mbps (similar to the \proack's resulted
throughput). The CWND growth for \proack and regular TCP shares a
similar pattern, which implies that our design and implementation can
efficiently and transparently aggregate multiple slow uplinks.

\subsubsection{Impact of network latency}
For TCP transmissions, RTT is an important factor that has impact on
the throughput.  In another experiment, we measure the performance of
\mytit with 4 different network latency settings listed in
Table~\ref{tab:latency}.  Each latency represents certain application
scenarios. For example, when users upload HD video with \mytit to CDN
edge servers, the latency to CDN servers is generally the regional
latency (32ms). When users upload data to their friends in another
continent, the RTT is on average 192ms.  Besides, according to our
measurements in a residential WiFi testbed~\cite{open-infrastructure},
we observe that the latency between broadband subscribers may vary
considerably, ranging between 20ms and 80ms. This depends on whether
users are with the same or a different ISP. Consider the case that a
user uploads data with \mytit through neighboring APs to another user
in the same city, the latency between the \ap{}s and the end user can
be quite different.

Given a certain number of APs, we assign to each \ap a random RTT
value between 20ms and 80ms. We carry out this test for 10 runs and
report the average throughput. As shown in
Figure~\ref{fig:latency-pro}, \proack throughput slightly declines as
network latency increases. In random latency setting, the resulted
throughput shows no significant difference.

\subsubsection{Impact of lossy wireless links}
%To have a rough idea of the reception performance of overhearing on monitor-APs, 
%we place one sender and one home-AP in the lab, with 5 meter distance.  We placed 
The wireless links in a real neighbourhood can be very lossy for a
variety of reasons, such as cross channel interference and distant
neighboring APs. Besides, since \mon{}s switch between transmit and
receive mode, they cannot overhear all transmitted packets.  To
estimate the potential of \mytit highly lossy wireless environments,
we emulate packet loss at \mon{}s by dropping received packets with a
probability $P$.  No losses were inflicted on \home, because \s
carries out unicast to \home, and 802.11 MAC already handles packet
loss and retransmissions automatically. We conduct the experiment with
3 values of $P$: 20\%, 40\%, and 60\%.

As indicated in Figure~\ref{fig:diversity-pro}, the throughput
reduction on lossy wireless links is very limited in all cases. The
good performance can be explained by the link diversity combined with
the centralized scheduling mechanisms.  The probability of some packet
not overheard by \emph{at least one} \mon is negligible small,
especially in case of high number of participating APs. This also
explains why 7~\ap{}s achieve higher throughput with $P=60\%$ than
with $P=20\%$.
%For higher $P$, 
%there is still a good chance that some \mon overhears the packet and shares the traffic
%load from home-AP. Whereas, less packets are reported to the BaPu-Gateway. The 
%reduced control overhead therefore leads to higher effective throughput.

\begin{figure}
\centering
\includegraphics[width=0.9\columnwidth]{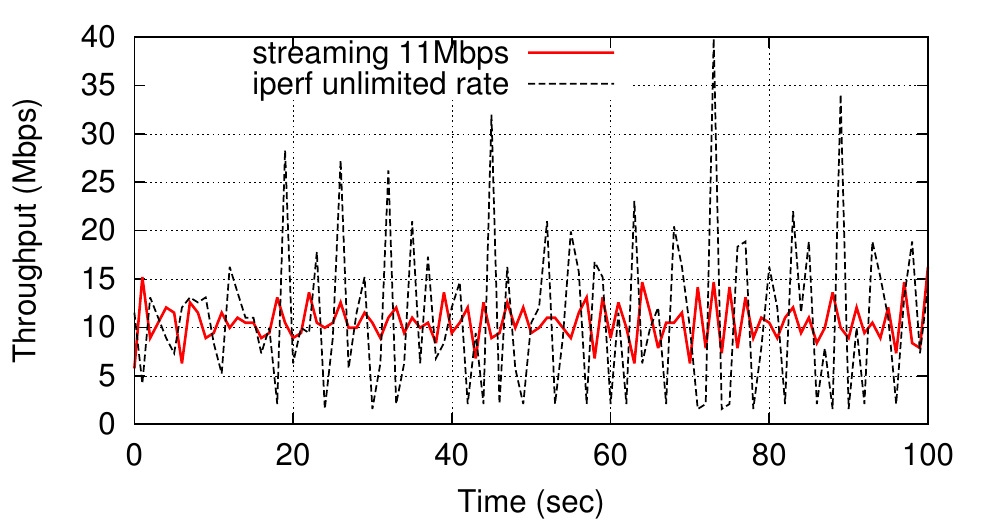}
\caption{Instantaneously received throughput comparison: 11Mbps
  Streaming vs. Unlimited rate.}
\label{fig:bursty}
\end{figure}

\subsubsection{Streaming vs. large file transfer}
One important goal in \mytit's design is to support instant sharing of
high-bitrate HD videos directly between users using streaming.  The
motivation behind is that today the major online streaming services
(e.g., Netflix) run on TCP based streaming technologies, such as HTTP
based Adaptive Bitrate Streaming.  Real time streaming generally
requires \emph{stable} instantaneous throughput.  In this experiment,
we study the potential of \mytit as a solution to high-bitrate
real-time streaming.

To emulate the streaming traffic, we use {\tt nuttcp} to issue a TCP
flow with a fixed 11Mbps sending rate. Figure~\ref{fig:bursty} shows
\d's instantaneous throughput in a 100 second session. \d achieves
a reasonably stable throughput in the whole session. It indicates that
\mytit can sustain high-bitrate streaming through aggregated
uplinks. In comparison, the {\tt iperf} flow with unlimited sending
rate shows much higher fluctuation.

\ignore{
During our {\tt iperf} measurement, even though the average TCP throughput is quite stable in various settings, we find that 
the \d's instantaneous throughput fluctuates considerably (Figure~\ref{fig:bursty}). 
Our packet trace inspection reveals that due to the latency difference
among APs, the arrival order of the scheduled TCP segments is uncertain. Therefore, the BaPu-Gateway sometimes must buffers 
the out-of-order segments until the expected ones arrive. Besides, since iperf always tries to saturate the link, as the throughput on the WiFi link overruns the aggregated backhual uplink capacity, the out-of-order segment arrival becomes more severe. Both reasons result in the bursty receiver end throughput. 

The iperf throughput only indicates that BaPu is suitable for some applications like instant backup of large files in the cloud. This aligns with the findings of prior work. However, it tells only one side of the story. As we design BaPu, the other important goal is to support instant sharing of high bitrate HD video directly from users' home WiFi, in streaming mode.  The motivation behind such goal is that today's main stream online streaming services (e.g. Netflix) run on TCP based streaming technologies, such as HTTP based Adaptive Bitrate Streaming.  Real time streaming generally requires stable instantaneous throughput.  In this experiment, we would like to study the potential of BaPu as a solution to high bitrate real time streaming. 

Unlike file uploading, streaming application generally has fixed transmitter rate, determined by the codec bit rate. To emulate the streaming traffic, we issue the TCP flow with a tool called \emph{nuttcp}. nuttcp can do rate limiting at transmitter. Figure \ref{fig:bursty} shows the receiver end instantaneous throughput every second in a 100 second session. In the streaming flow with 11Mbps fixed transmitter rate, the receiver end achieves reasonably stable throughput in the whole session. It indicates that BaPu can sustain high bit rate streaming through aggregated links. In comparison, the iperf flow with unlimited transmitter rate shows much higher fluctuation. }

\section{Impact of Uplink Sharing}
\label{sec:uplink}

\begin{figure}
\centering
\includegraphics[width=0.9\linewidth]{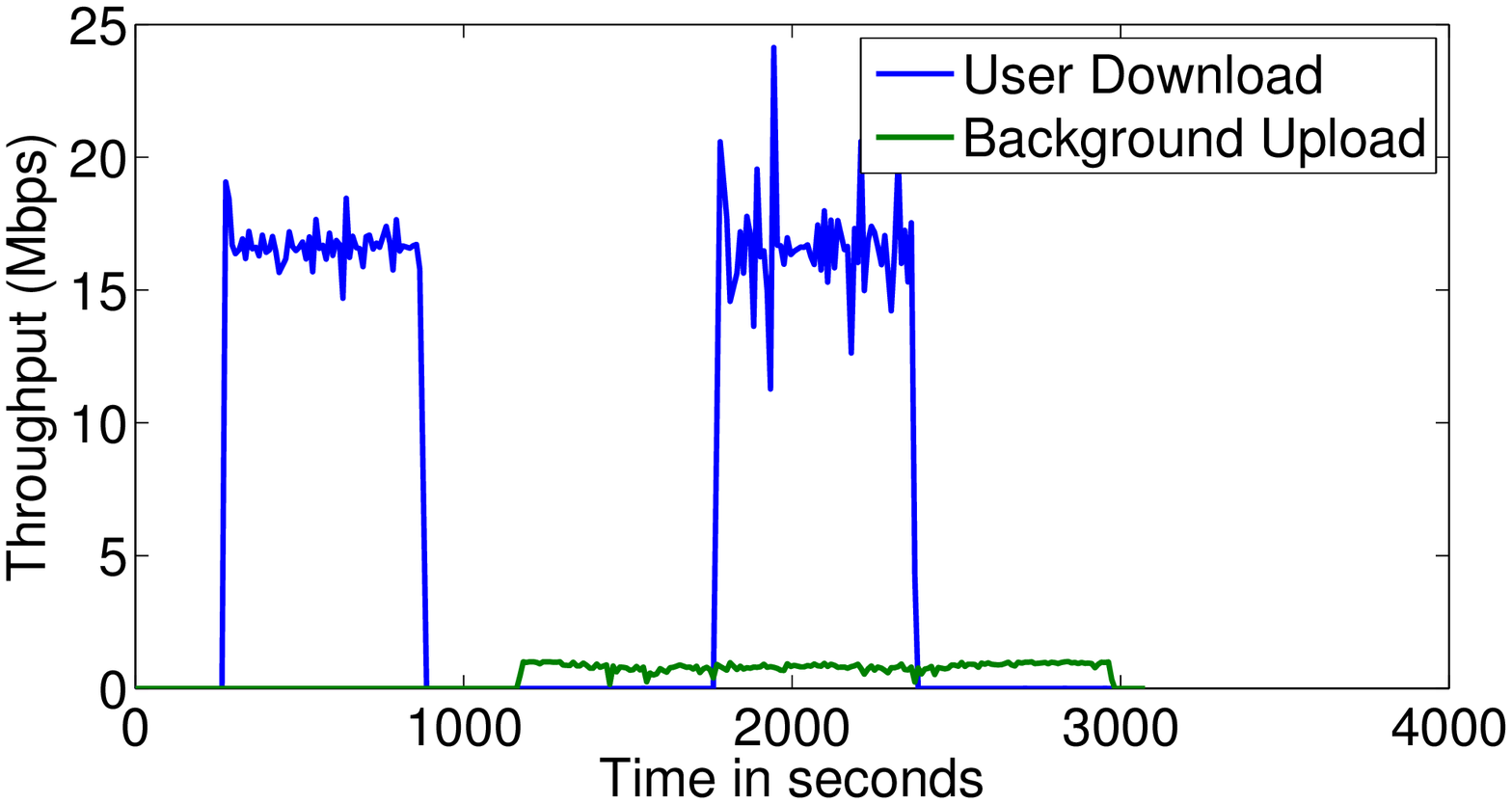}
\caption{Regular user download behaviour with and without background upload traffic. No impact is
apparent in the presence of competing background traffic.}
\label{fig:dl-backoff}
\end{figure}

\mytit is essentially a crowd-sourcing approach that shares users' idle bandwidth to help others.  The goal is to harness as much idle bandwidth as possible with minimal impact on home users' network usage.  We first show that with standard Linux traffic shaping tools, bandwidth sharing has minimal impact on regular traffic. Next, we study how much bandwidth can be harnessed with a testbed in residential broadband.

\subsection{Prioritizing Home User's Traffic}

Techniques and tools for traffic shaping, such as Linux's {\tt tc} are widely available. While a full-fledged system may use complicated traffic prioritization, it is sufficient for \mytit to simply classify traffic in two classes: regular user traffic with the highest priority, and background sharing traffic with a minimum bandwidth guarantee, but allowed to grow if more bandwidth is available. To implement this, we use \textit{Hierarchical Token Bucket} and
\textit{Stochastic Fair Queuing} modules within {\tt tc}  to fairly distribute traffic belonging to
the same class.

\begin{figure}
\centering
\includegraphics[width=0.9\linewidth]{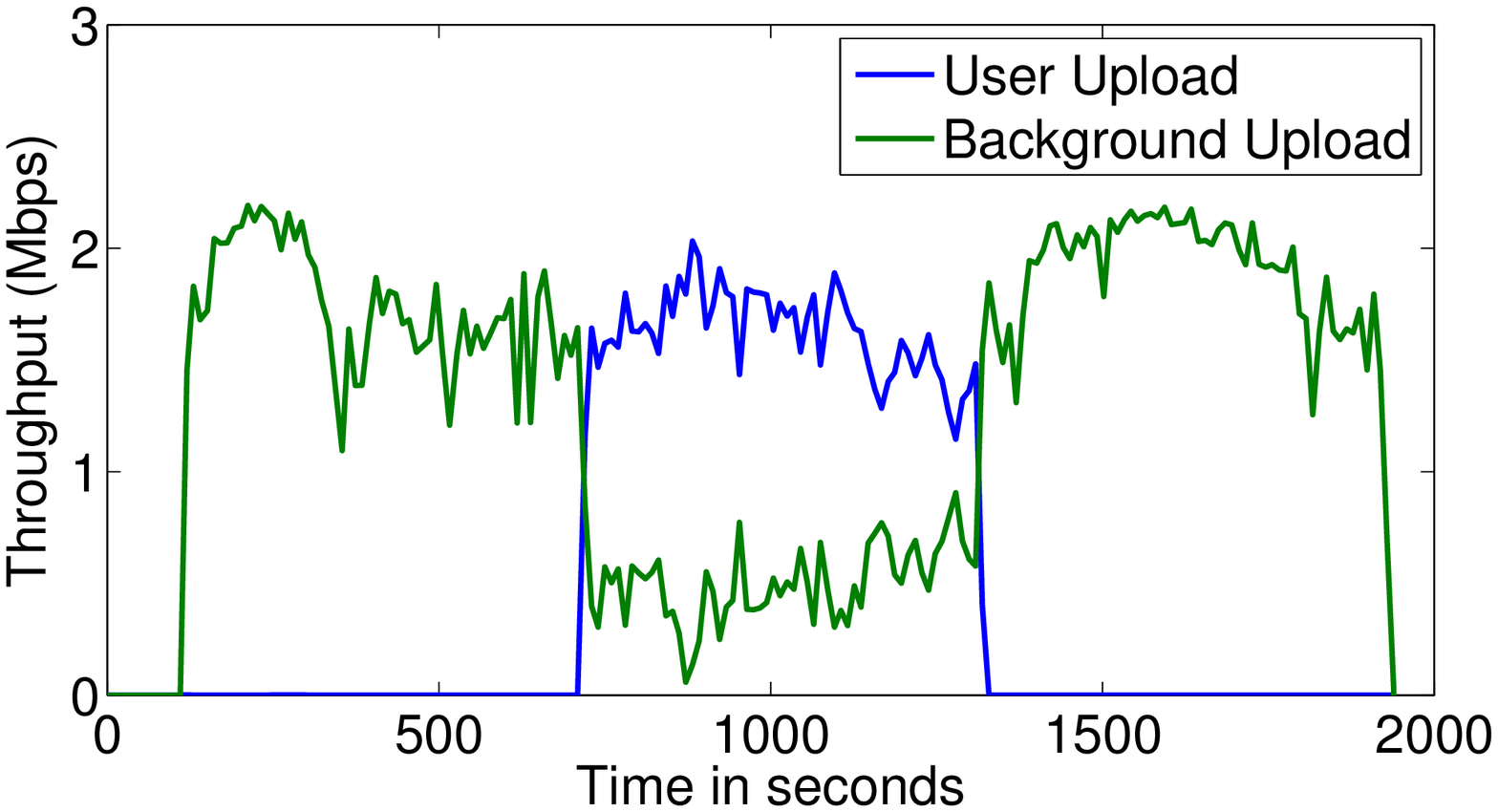}
\caption{Regular uplink observation with and without background traffic.
The low-priority background traffic backs off as soon as regular traffic starts.}
\label{fig:ul-backoff}
\end{figure}

We set up the traffic shaping on one OpenWRT home router, and validate the correctness of the traffic shaping with two tests.  In the first test, we first 
generate regular download traffic for 10 minutes to obtain a baseline of AP's download throughput. After the baseline measurement, we start the background upload for 20 minutes, emulating the uplink sharing. As the upload goes on, we relaunch the regular download traffic.  As shown in Figure~\ref{fig:dl-backoff}, the user's regular download throughput is not affected, because the TCP ACKs related to the user download have been prioritized. Also, TCP ACKs consume limited
uplink bandwidth, and thus has negligible impact to the background upload. In the second test, we examine the analogous case for regular upload traffic. We first start emulated background upload with a minimum bandwidth guarantee of 500Kbps. During the background upload, we start the regular user upload.  As shown in Figure~\ref{fig:ul-backoff}, the regular upload traffic takes over more bandwidth, while the background upload backs off properly (but not lower than 500Kbps). We conclude that with proper traffic shaping, \mytit can provide uplink sharing with a minimal impact on users' regular traffic.

\subsection{Push Uplink Sharing to the Limit}

To find out how much idle uplink bandwidth can be harnessed, we instrument an uplink throughput experiment to 17 residential APs, covering 2 major ISPs in Boston (Table~\ref{table:ul_exp_summary}). Each AP is configured with proper traffic shaping. With constant background upload, each AP reports background upload throughput every 10 seconds, lasting for 16 days.

\begin{table}[ht]
\caption{Uplink experiment data summary}
\label{table:ul_exp_summary}
\centering
\begin{tabular}{lc}
\hline
\textbf{Home APs}&Comcast (14), RCN (3)\\
\textbf{Data collection time}& May 22 -- Jun 13, 2012\\
\textbf{Mean AP online time}&381 hours ($\sim$ 16 days)\\
\textbf{Throughput samples}&2.3 millions\\%2,330,580\\
\hline
\end{tabular}
\end{table}

\begin{figure}
\center
\includegraphics[width=0.9\linewidth]{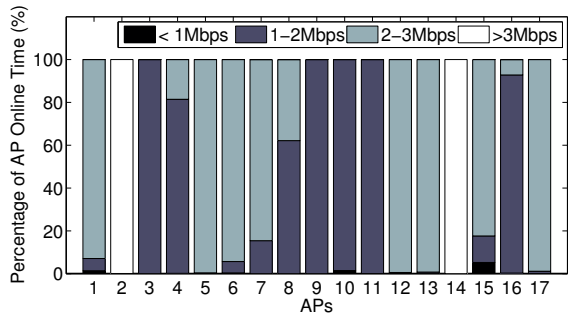}
\caption{Fraction of time spent by each AP with certain background upload throughput over the AP's total online time.}
\label{fig:exp_ul_bw_breadown_per_ap}
\end{figure}

\begin{figure}
\centering
\includegraphics[width=0.9\linewidth]{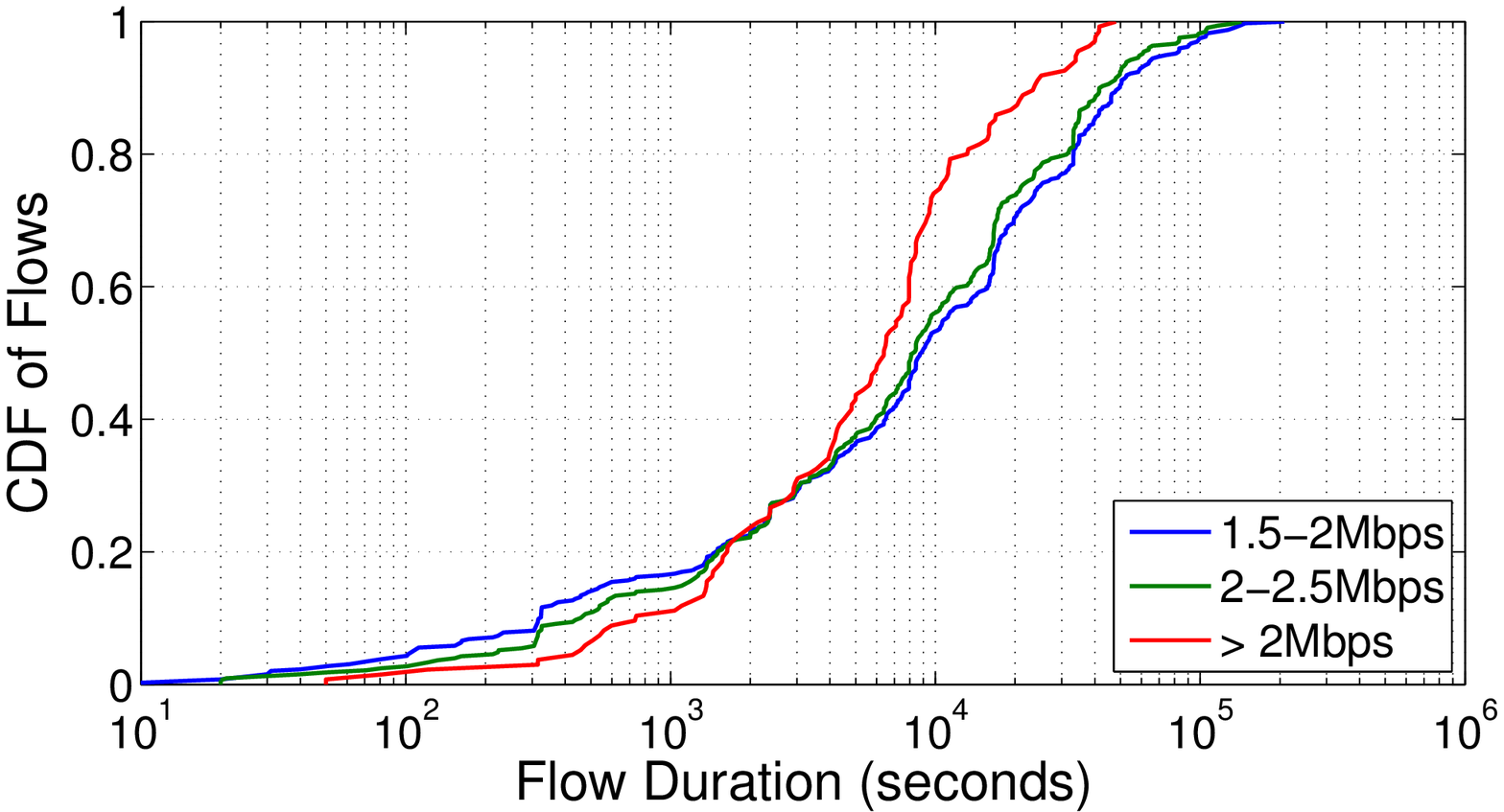}
\caption{Duration of background traffic flow of certain throughput classes.}
\label{fig:exp_ul_flow_length_cdf-peakhour}
\end{figure}

As shown in Figure~\ref{fig:exp_ul_bw_breadown_per_ap}, all APs can share 1-3Mbps uplink throughput during most of the experiment time.  Each AP's mean shared throughput is very close to its uplink limit set by ISP.  Also, we investigate how long a certain upload throughput can last. Figure~\ref{fig:exp_ul_flow_length_cdf-peakhour} shows the the CDF of duration for which the background upload can sustain certain throughput during peak hours (18pm$\sim$23pm).  We see that, there is over 80\%  chance that the background upload throughput can stay above 2Mbps for over 30 minutes. To conclude, there is abundant idle uplink bandwidth in residential broadband that can be exploited for upload sharing.

\section{Related Work}
\label{sec:related}

% Grouping Wireless APs

\mytit system is inspired by design principles of several earlier protocols, it however addresses unique constraints and goals and results in a novel combination of techniques that achieves high efficiency. Several earlier research works proposed to improve the performance of TCP over wireless links by using intermediate nodes to assist in the recovery of lost packets include Snoop TCP~\cite{BalakrishnanSAK1995}, and  Split TCP for ad hoc networks~\cite{KoppartyKFT2002}. Multiple radio links to improve devices throughput have also been explored
from several perspectives including traffic aggregation~\cite{KandulaLBK2008}, mutipath
forwarding~\cite{link-alike}, mitigation of wireless losses~\cite{MiuBK2005,miu2004divert}. In
addition to systems that rely on multiple radio interfaces~\cite{BahlAPW2004},
many solutions and algorithms were proposed for a single radio interface that
carefully switches across multiple access points while providing the upper
layers of the network stack a transparent access through a virtual
interface~\cite{VirtualWiFi,ChandraB2004,XingML2010, KandulaLBK2008}. Solutions to overcome
the limited APs backhaul through aggregation using a virtualized radio interface
include the initial Virtual-WiFi system where two TCP connection might be
services by through two different APs~\cite{VirtualWiFi}, FatVAP that achieves
fast switching a smart AP selection~\cite{KandulaLBK2008}, ARBOR that add
security support~\cite{XingML2010}, Fair WLAN that provides
fairness~\cite{GiustinianoGTLDMR2010}. Many of these systems require techniques
for fast switching across access points to reduce the impact on TCP performance
in terms of delay and packet loss as proposed in Juggler~\cite{NicholsonWN2009}, and 
WiSwitcher~\cite{GiustinianoGLR2009}. In \cite{Soroush:2011ki}, an analytical model 
is proposed to optimze concurrent AP connections for highly mobile clients.  They also implement
Spider, a multi-AP driver using optimal AP and channel scheduling to improve the aggregated throughput. 
Unlike BaPu, these works do not aggregate the throughput for single transport layer connection, which is critical for client transparency.
Divert~\cite{miu2004divert} propose a central controller to select the optimal AP across multiple BSes
in WLANs in order to reduce the path-dependent downlink loss from AP to client.  
Rather than improving the wireless link quality, BaPu is aimed to aggregate the wired capacity behind APs. 
In BaPu, the sender communicates with its home AP as usual.  
However, BaPu does benefit from the link diversity across APs while aggregating the backhaul capacity.
ViFi~\cite{Balasubramanian:2008vo} propose a probabilistic algorithm for coordination between basestations to 
improve the opportunistic connectivity of the client-BS communication. Similar to Divert, ViFi is not for aggregating throughput. Also, in section~\ref{sec:schedule}, we show that such probabilistic solution sets limitations on throughput aggregation. 
The closest approach to our work is the Link-alike system where
access points coordinate to opportunistically schedule the traffic over their backhaul links~\cite{link-alike}. 
Our approach differs from previous work in requiring that the client devices remain 
unchanged (unicast connection to the AP) and transparently supports protocols like TCP. 

Being completely transparent to the clients and constraining each link
AP-Destination flow to be TCP-friendly makes efficient multipath transport, a
key component of our system. There has been a significant amount of work in this
area for quite some time from various perspectives that are fairly different
from our setup. Previous work identified the issues with differential delay, and
rates and mostly focussed on providing reliability, flows balancing, and
maintaining fairness. Proposed solutions, require the modification of
the client devices network stacks, and usually do not aim at increasing capacity
through simultaneous use of all paths. For example, the IETF has two standards
on transport protocols using multipath. The Stream Control Transmission Protocol
(SCTP) was primarily designed for multi-homed devices that require fail-over
support~\cite{SCTP}, the more recent Multi-Path TCP (MPTCP) is a TCP extension
that aims at enabling nodes to efficiently communicate utilizing multiple
parallel paths~\cite{MPTCP}. In recent work,~\cite{WischikRGH2011} proposed a
congestion control mechanism for MPTCP with the objective of reliability and
fairness. Other transport protocols that require the modification of the client
devices, include pTCP~\cite{HsiehS2002} an end to end transport protocol that
achieves bandwidth aggregation on multi-homed mobile hosts,
RCP~\cite{HsiehKZS2003} a Receiver Centric Protocol that aggregates
heterogeneous interfaces traffic and carries link specific congestion control,
R-MTP~\cite{MagalhaesK2001} balances and coordinates the traffic over wireless
links with varying characteristics, Horizon~\cite{RadunovicGG2008} uses
back-pressure technique to balance the traffic across forwarding paths. Beyond
mobile communication environments, multipath TCP features have also been finding
applications in various networking environments such as data
~\cite{RaiciuBPGW2011, BarrePB2011}. The distinguishing element of \mytit is that 
it aims at transparently supporting unmodified client devices and TCP/IP stacks while efficiently 
aggregating the APs backhauls. 

% \fixme{XXX Relays:SoftRepeater}

\section{Conclusion}
\label{sec:conclusion}
In this work, we present the design and implementation of BaPu, a complete software based solution on WiFi APs for aggregating multiple broadband uplinks.  First, with a large scale wardriving data and a long term measurement in Boston residential broadband, we show that the high AP density and under utilized broadband uplinks calls for a solution to harness the idle bandwidth for a boosted uplink throughput. 

With our client transparent design, BaPu offers generic support for legacy client and a large variety of network applications.  However, the client transparency design raises many new technical challenges. In particular, we propose a novel mechanism, called Proactive-ACK, to address the challenge of multiplexing single TCP session through multiple paths without degrading performance. The benefit of such mechanism is analysed with experimental data. We carry out an extensive set of experiments to evaluate the BaPu throughput performance for both UDP and TCP, in a variety of realistic network settings.  BaPu achieves over 95\% aggregation efficiency for UDP and over 88\% for TCP, even in lossy wireless environment.  

Besides, to further justify the feasibility of BaPu as a crowd-sourcing mechanism, we empirically show the potential idle uplink bandwidth that can be harnessed from residential broadband networks.  We also provide a design guideline for such bandwidth sharing system to eliminate the negative impact to home users. Also, the software based solution makes BaPu an easy incremental deployment, especially as APs are becoming social and cloud-managed. 

\bibliography{bapu_mobisys}

\begin{thebibliography}{31}
\providecommand{\natexlab}[1]{#1}
\providecommand{\url}[1]{\texttt{#1}}
\expandafter\ifx\csname urlstyle\endcsname\relax
  \providecommand{\doi}[1]{doi: #1}\else
  \providecommand{\doi}{doi: \begingroup \urlstyle{rm}\Url}\fi

\bibitem[aka(2011)]{akamai:hd}
{Akamai HD Network}.
\newblock \emph{Akaimai Technical Report}, 2011.
\newblock
  \\\url{http://www.akamai.com/dl/brochures/Akamai_HD_Network_Solution_Brochure.pdf}.

\bibitem[Akella et~al.(2007)Akella, Judd, Seshan, and
  Steenkiste]{AkellaJSS2007}
A.~Akella, G.~Judd, S.~Seshan, and P.~Steenkiste.
\newblock Self-management in chaotic wireless deployments.
\newblock \emph{Wireless Networks}, 2007.

\bibitem[Allman et~al.(2009)Allman, Paxson, and
  Blanton]{Allman:2009:TCC:RFC5681}
M.~Allman, V.~Paxson, and E.~Blanton.
\newblock Tcp congestion control, 2009.

\bibitem[Bahl et~al.(2004)Bahl, Adya, Padhye, and Walman]{BahlAPW2004}
P.~Bahl, A.~Adya, J.~Padhye, and A.~Walman.
\newblock {Reconsidering wireless systems with multiple radios}.
\newblock \emph{SIGCOMM Compututer Communication Review}, 34:\penalty0 39--46,
  2004.
\newblock {ISSN} 0146-4833.

\bibitem[Balakrishnan et~al.(1995)Balakrishnan, Seshan, Amir, and
  Katz]{BalakrishnanSAK1995}
Hari Balakrishnan, Srinivasan Seshan, Elan Amir, and Randy~H. Katz.
\newblock {Improving TCP/IP performance over wireless networks}.
\newblock In \emph{Proceedings of MobiCom}, 1995.

\bibitem[Balasubramanian et~al.(2008)Balasubramanian, Mahajan, Venkataramani,
  Levine, and Zahorjan]{Balasubramanian:2008vo}
A~Balasubramanian, R~Mahajan, A~Venkataramani, BN~Levine, and J~Zahorjan.
\newblock {Interactive wifi connectivity for moving vehicles}.
\newblock \emph{Proc. of SigComm}, 2008.

\bibitem[Barr{\'e} et~al.(2011)Barr{\'e}, Paasch, and Bonaventure]{BarrePB2011}
S.~Barr{\'e}, C.~Paasch, and O.~Bonaventure.
\newblock {MultiPath TCP: from theory to practice}.
\newblock In \emph{Proceedings of NETWORKING}, 2011.

\bibitem[Chandra and Bahl(2004)]{ChandraB2004}
R.~Chandra and P.~Bahl.
\newblock Multinet: connecting to multiple ieee 802.11 networks using a single
  wireless card.
\newblock In \emph{Proc. of INFOCOM}, 2004.

\bibitem[FON(2012)]{fon}
FON.
\newblock {FON}, 2012.
\newblock \\\url{http://corp.fon.com/us/}.

\bibitem[Ford et~al.(2012)Ford, Raiciu, Handley, and Bonaventure]{MPTCP}
A.~Ford, C.~Raiciu, M.~Handley, and O.~Bonaventure.
\newblock {TCP Extensions for Multipath Operation with Multiple Addresses}.
\newblock Internet-Draft, 2012.

\bibitem[Giustiniano et~al.(2009)Giustiniano, Goma, Lopez, and
  Rodriguez]{GiustinianoGLR2009}
D.~Giustiniano, E.~Goma, A.~Lopez, and P.~Rodriguez.
\newblock Wiswitcher: an efficient client for managing multiple aps.
\newblock In \emph{Proceedings of PRESTO}, 2009.

\bibitem[Giustiniano et~al.(2010)Giustiniano, Goma, Toledo, Dangerfield,
  Morillo, and Rodriguez]{GiustinianoGTLDMR2010}
D.~Giustiniano, E.~Goma, A.L. Toledo, I.~Dangerfield, J.~Morillo, and
  P.~Rodriguez.
\newblock {Fair WLAN backhaul aggregation}.
\newblock In \emph{MobiCom}, 2010.

\bibitem[Hsieh and Sivakumar(2002)]{HsiehS2002}
H.-Y. Hsieh and R.~Sivakumar.
\newblock {A transport layer approach for achieving aggregate bandwidths on
  multi-homed mobile hosts}.
\newblock In \emph{Proceedings of MobiCom}, 2002.

\bibitem[Hsieh et~al.(2003)Hsieh, Kim, Zhu, and Sivakumar]{HsiehKZS2003}
H.-Y. Hsieh, K.-H. Kim, Y.~Zhu, and R.~Sivakumar.
\newblock {A receiver-centric transport protocol for mobile hosts with
  heterogeneous wireless interfaces}.
\newblock In \emph{Proceedings of MobiCom}, 2003.

\bibitem[Jakubczak et~al.(2008)Jakubczak, Andersen, Kaminsky, Papagiannaki, and
  Seshan]{link-alike}
S.~Jakubczak, D.~G. Andersen, M.~Kaminsky, K.~Papagiannaki, and S.~Seshan.
\newblock {Link-alike: using wireless to share network resources in a
  neighborhood}.
\newblock \emph{SIGMOBILE Mobile Computing Communications Review}, 2008.

\bibitem[Kandula et~al.(2008)Kandula, Lin, Badirkhanli, and
  Katabi]{KandulaLBK2008}
S.~Kandula, K.C. Lin, T.~Badirkhanli, and D.~Katabi.
\newblock {FatVAP: Aggregating AP backhaul capacity to maximize throughput}.
\newblock In \emph{Proceedings of NSDI}, 2008.

\bibitem[Kopparty et~al.(2002)Kopparty, Krishnamurthy, Faloutsos, and
  Tripathi]{KoppartyKFT2002}
Swastik Kopparty, Srikanth~V. Krishnamurthy, Michalis Faloutsos, and Satish~K.
  Tripathi.
\newblock Split tcp for mobile ad hoc networks.
\newblock In \emph{GLOBECOM}, 2002.

\bibitem[Magalhaes and Kravets(2001)]{MagalhaesK2001}
L.~Magalhaes and R.H. Kravets.
\newblock {Transport Level Mechanisms for Bandwidth Aggregation on Mobile
  Hosts}.
\newblock In \emph{Proceedings of Conference on Network Protocols}, 2001.

\bibitem[Meraki(2012)]{meraki}
Meraki.
\newblock {Meraki}, 2012.
\newblock \\\url{http://www.meraki.com/}.

\bibitem[{Microsoft Research}(2012)]{VirtualWiFi}
{Microsoft Research}.
\newblock Virtual wifi, 2012.
\newblock
  \\\url{http://research.microsoft.com/en-us/um/redmond/projects/virtualwifi/}.

\bibitem[Miu et~al.()Miu, Balakrishnan, and Koksal]{MiuBK2005}
Allen Miu, Hari Balakrishnan, and Can~Emre Koksal.
\newblock Improving loss resilience with multi-radio diversity in wireless
  networks.
\newblock MobiCom '05.

\bibitem[Miu et~al.(2004)Miu, Tan, Balakrishnan, and
  Apostolopoulos]{miu2004divert}
Allen~K. Miu, Godfrey Tan, Hari Balakrishnan, and John Apostolopoulos.
\newblock Divert: Fine-grained path selection for wireless lans.
\newblock In \emph{Proc. of MobiSys}, 2004.

\bibitem[Nicholson et~al.(2010)Nicholson, Wolchok, and Noble]{NicholsonWN2009}
Anthony~J. Nicholson, Scott Wolchok, and Brian~D. Noble.
\newblock Juggler: Virtual networks for fun and profit.
\newblock \emph{IEEE Transactions on Mobile Computing}, 9:\penalty0 31--43,
  2010.

\bibitem[Open Infrastructure()]{open-infrastructure}
Open Infrastructure.
\newblock Open infrastructure: A wireless network research framework for
  residential networks.
\newblock
  \\\url{http://www.ccs.neu.edu/home/noubir/projects/openinfrastructure/},
  2012.

\bibitem[OpenWRT(2012)]{openwrt}
OpenWRT.
\newblock {OpenWRT -- Wireless Freedom}, 2012.
\newblock \\\url{http://openwrt.org/}.

\bibitem[Radunovi\'{c} et~al.(2008)Radunovi\'{c}, Gkantsidis, Gunawardena, and
  Key]{RadunovicGG2008}
B.~Radunovi\'{c}, C.~Gkantsidis, D.~Gunawardena, and P.~Key.
\newblock {Horizon: Balancing TCP over multiple paths in wireless mesh
  network}.
\newblock In \emph{Proc. of MobiCom}, 2008.

\bibitem[Raiciu et~al.(2011)Raiciu, Barre, Pluntke, Greenhalgh, Wischik, and
  Handley]{RaiciuBPGW2011}
C.~Raiciu, S.~Barre, C.~Pluntke, A.~Greenhalgh, D.~Wischik, and Ma. Handley.
\newblock {Improving datacenter performance and robustness with multipath TCP}.
\newblock In \emph{SIGCOMM '11}, 2011.

\bibitem[Soroush et~al.(2011)Soroush, Gilbert, Banerjee, Levine, Corner, and
  Cox]{Soroush:2011ki}
Hamed Soroush, Peter Gilbert, Nilanjan Banerjee, Brian~Neil Levine, Mark
  Corner, and Landon Cox.
\newblock {Concurrent Wi-Fi for mobile users: analysis and measurements}.
\newblock In \emph{CoNEXT '11}, 2011.

\bibitem[Steward(2007)]{SCTP}
R.~Steward.
\newblock Stream control transmission protocol.
\newblock IETF RFC 4960, 2007.

\bibitem[Wischik et~al.(2011)Wischik, Raiciu, Greenhalgh, and
  Handley]{WischikRGH2011}
D.~Wischik, C.~Raiciu, A.~Greenhalgh, and M.~Handley.
\newblock {Design, implementation and evaluation of congestion control for
  multipath TCP}.
\newblock In \emph{Proc. of NSDI}, 2011.

\bibitem[Xing et~al.(2010)Xing, Mishra, and Liu]{XingML2010}
X.~Xing, S.~Mishra, and X.~Liu.
\newblock {ARBOR: hang together rather than hang separately in 802.11 wifi
  networks}.
\newblock In \emph{Proc. of INFOOCM}, 2010.

\end{thebibliography}

\end{document}